\documentclass[journal,draftclsnofoot,onecolumn,12pt]{IEEEtran}%

\usepackage{amsmath}
\usepackage{tikz}

\usepackage{amssymb}

\usepackage{tabu}
\usepackage{comment}
\usepackage{todonotes}
\usepackage{epsf}
\usepackage{epsfig}
\usepackage{times}
\usepackage{epsfig}
\usepackage{graphicx}
\usepackage{bbold}
\usepackage{mathtools}
\usepackage{mathrsfs}
\usepackage{amssymb,amsmath}
\usepackage{pdfpages}
\usepackage{epstopdf}

\usepackage{dsfont}
\usepackage{lettrine} 

\usepackage{amsmath,epsfig,amssymb,amsthm,cite,url}

\usepackage[justification=centering]{caption}
\usepackage{cite}
\usepackage{bbm}
\usepackage{algorithmic}
\usepackage{algorithm}

\allowdisplaybreaks
\usepackage{csquotes}
\usepackage{amsmath,amssymb}

\usepackage{verbatim}
\usepackage[english]{babel}
\usepackage{amsmath,amssymb}

\usepackage{verbatim}

\newtheorem{corollary}{Corollary}
\let\oldcorollary\corollary
\renewcommand{\corollary}{\oldcorollary\normalfont}
\newtheorem{theorem}{\bf Theorem}
\let\oldtheorem\theorem
\renewcommand{\theorem}{\oldtheorem\normalfont}

\newtheorem{proposition}{\bf Proposition}
\let\oldproposition\proposition
\renewcommand{\proposition}{\oldproposition\normalfont}
\newtheorem{lemma}{\bf Lemma}
\let\oldlemma\lemma
\renewcommand{\lemma}{\oldlemma\normalfont}
\newtheorem{example}{Example}
\let\oldexample\example
\renewcommand{\example}{\oldexample\normalfont}

\newtheorem{definition}{\bf Definition}
\let\olddefinition\definition
\renewcommand{\definition}{\olddefinition\normalfont}
\newtheorem{remark}{Remark}
\let\oldremark\remark
\renewcommand{\remark}{\oldremark\normalfont}

\usepackage{changes}
\definechangesauthor[name={Ali Zadeh}, color=blue]{ali}
\setremarkmarkup{(#2)}

\usepackage{color}
\usepackage{xcolor}
\usepackage{dsfont}
\newcommand{\mub}{\boldsymbol{\mu}}

\newcommand{\Sigmab}{\boldsymbol{\Sigma}}
\newcommand{\rhob}{\boldsymbol{\rho}}
\newcommand{\Pb}{\boldsymbol{P}}
\newcommand{\xb}{\boldsymbol{x}}
\newcommand{\Xb}{\boldsymbol{X}}
\newcommand{\yb}{\boldsymbol{y}}
\newcommand{\Db}{\mathcal{D}}
\newcommand{\Ab}{\mathcal{A}}
\newcommand{\Eb}{\mathcal{E}}

\newcommand{\deltab}{\boldsymbol{\delta}}
\newcommand{\lambdab}{\boldsymbol{\lambda}}
\newcommand{\eb}{\mathbf{e}}
\newcommand{\wb}{\boldsymbol{w}}

\usepackage{tikz}
\usetikzlibrary{arrows}
\allowdisplaybreaks
\newcommand{\zb}{\boldsymbol{z}}

\usepackage{setspace}	

\begin{document}
	\newcommand{\bb}[1]{\mathbb{#1}}
	\newcommand{\mc}[1]{\mathcal{#1}}
	%
	%

	\title{Human-in-the-Loop Wireless Communications: Machine Learning and Brain-Aware Resource Management}

\author{\IEEEauthorblockN{  Ali Taleb Zadeh Kasgari$^1$, Walid Saad$^1$, and M\'erouane Debbah$^2$}\\
	\IEEEauthorblockA{
		\small $^1$ Wireless@VT, Electrical and Computer Engineering Department, Virginia Tech, VA, USA,\\ Emails:\url{{alitk , walids}@vt.edu}.\\
		$^2$ Mathematical and Algorithmic Sciences Lab, Huawei France R\&D, Paris, France, and CentraleSup´elec,\\   Universit´e Paris-Saclay, Gif-sur-Yvette, France, Email: \url{merouane.debbah@huawei.com}.
        \thanks{A preliminary version of this work appeared in the proceedings of the
51th Asilomar Conference on Signals, Systems and Computers, Pacific Grove, CA, USA \cite{Kasgari2017Asilomar}.}
\thanks{This research was supported by the U.S. National Science Foundation under Grants CNS-1460316 and IIS-1633363.}
	}}



	\maketitle
	
	\begin{abstract}
		Human-centric applications such as virtual reality
		and immersive gaming are central to future wireless networks.
		Common features of such services include: a) their dependence
		on the human user's behavior and state, and b) their need for more network resources compared to 	   conventional applications. To successfully
		deploy such applications over wireless networks, the
		network must be made cognizant of not only the quality-of-service (QoS) needs of the applications, but also of the perceptions of the \emph{human users} on this QoS. 
		In this paper, by explicitly modeling the limitations of the human brain, a concrete measure for the delay perception of human users is introduced. {Then, a  learning method, called probability distribution identification, is developed to find a probabilistic model for this delay perception based on the brain features of a human user.} 
        Given the learned model for the delay perception of the human brain, a brain-aware resource management algorithm based on Lyapunov optimization is proposed for allocating radio resources to human users while minimizing the transmit power and taking into account the reliability of both machine type devices and human users. Then, a closed-form relationship between the reliability measure and wireless physical layer metrics of the network is derived. Simulation results show that a brain-aware approach can yield savings of up to $78\%$ in power compared to the system that only considers QoS metrics. The results also show that, compared with QoS-aware, brain-unaware systems, the brain-aware approach can save substantially more power in low-latency systems.
	\end{abstract}


	%
	\IEEEpeerreviewmaketitle

	\section{Introduction}
	The next generation of wireless services is expected to be highly human centric. Examples include virtual reality and interactive/immersive gaming \cite{saad2019vision,gobbetti1998virtual,chen2017virtual}. 
    To cope with the quality-of-service (QoS) needs of such human-centric applications, in terms of data rate and ultra-low latency, wireless networks must exploit substantially more radio resources by leveraging heterogeneous spectrum bands  \cite{semiari2015context}. Although allocating heterogeneous spectrum resources can potentially increase the raw QoS, given the human-centric nature of emerging applications, their users may not be able to perceive the improved QoS, due to human factors such as the cognitive limitations of the brain \cite{RN1}. Indeed, many empirical studies (anecdotal and otherwise) have shown that the limitations on the human brain can be translated into a limitation on how wireless users translate QoS into actual quality-of-experience (QoE) \cite{Laghari2013Neuro,Wechsung2014Springer,Zhao2017QoE}. For example, the human brain may not be able to perceive any difference between videos transmitted with different QoS (e.g., rates or delays) \cite{Zhao2017QoE,Chen2015QoE}.

    Hence, in order to deploy these services over wireless networks, such as 5G cellular systems, there is a need to enable the system to be strongly cognizant of the human user in the loop. In particular, to deliver  immersive, human-centric services, the network must tailor the usage and optimization of wireless resources to the intrinsic features of its human users such as their behavior and brain processing limitations. By doing so, the network can potentially save resources, accommodate more users,  and provide a more realistic QoE to its users. {Moreover, the saved resources can be used to accommodate emerging applications in wireless networks such as drone communications \cite{rahmati2019dynamic,kasgari2019uav} and autonomous driving \cite{reviewer1Transportation,ferdowsi2018robust,reviewer5IndustInfo,reviewer2Network}}.

	Developing resource management mechanisms that can cater to intrinsic needs of wireless users and their context (e.g., device features or social metrics) has recently been studied in \cite{Alam2016,Lin2017TVT,Letter2017Context,semiari2015context,WCNC2017context,bennis2014proactive,makris2013survey,Yuen2011Location,proebster2011context}. In \cite{Alam2016}, a context-aware scheduling algorithm for 5G systems is proposed. This algorithm exploits the context information of user equipments (UEs), such as battery level, to save energy in the system while satisfying the QoS requirements of users. The authors in \cite{Lin2017TVT}, proposed a user-centric resource allocation framework for ultra-dense heterogeneous networks.
    Context-aware resource allocation for heterogeneous cellular networks is also studied in \cite{Letter2017Context,semiari2015context}, and \cite{WCNC2017context}.
	In \cite{semiari2015context}, a novel approach to context-aware resource allocation in small cell networks is introduced. Both wireless physical layer metrics and the social ties of human users are exploited in \cite{semiari2015context} to allocate wireless resource blocks.  
	Proactive caching  using context information from social networks is studied in \cite{bennis2014proactive}. The results in \cite{bennis2014proactive} show that such a socially-aware caching technique reduces the peak traffic in 5G networks. Other context-aware resource allocation algorithms are also studied in 
	\cite{makris2013survey,perera2014context}, and \cite{proebster2011context}.
	{However, despite this surge in literature on context-aware networking \cite{Alam2016,semiari2015context,bennis2014proactive,makris2013survey,Letter2017Context,WCNC2017context,Lin2017TVT,proebster2011context}, and \cite{perera2014context} this  prior art is still reliant on device-level features. Moreover, the works in \cite{Alam2016,semiari2015context,bennis2014proactive,makris2013survey,Letter2017Context,WCNC2017context,Lin2017TVT,proebster2011context}, and \cite{perera2014context} are agnostic to the human users and their features (e.g., brain limitation or behavior). Hence, adapting these existing approaches can waste network resources {due to the potential allocation of} more resources to human users that cannot perceive the associated QoS gains, due to cognitive brain limitations.}

	A general framework for modeling the intelligence of communication systems which serve humans is proposed in \cite{huang2016system}. The author defines  intelligence in terms of predicting  and serving human demands in advance.
	However, the work in \cite{huang2016system} does not account for the cognitive limitations of a human brain. Moreover,  demand prediction, as done in \cite{huang2016system}, will not be sufficient to capture the full spectrum of the human user limitations and behavior. 
    By being aware of brain limitations of each user, the  network can provide a unique experience for each user and optimize its performance.
	For example, an increase in the delay of a wireless system  may have different effects on the QoE perceived by different human users. 
	In particular, such different delay perceptions can potentially be exploited by the cellular network to minimize power consumption and reduce the amount of wasted resources. To our best knowledge, no existing work has studied the impact of such disparate brain delay perceptions on wireless resource allocation. 
	systems.
	{Furthermore, none of the prior studies on systems with humans-in-the-loop in other fields \cite{reviewer3Cybernetics} and \cite{reviewer4ImagePro} have analyzed the human brain limitations.}
	
	{The main contribution of the paper is a \emph{novel brain-aware learning and resource management framework} that  explicitly factors in the brain state of human users during resource allocation in a cellular network.}
	In particular, we  formulate the brain-aware resource allocation problem using a joint learning and optimization framework. {First, we propose a learning algorithm to identify the delay perceptions of a human brain.} This learning algorithm employs both supervised and unsupervised learning to identify the brain limitations and also creates a statistical model for these limitations based on Gaussian mixture models. Then, using Lyapunov optimization, we address the resource allocation problem with time varying QoS requirements that captures the learned delay perception. Using this approach, the network can allocate radio resources to human users while considering the reliability of both machine type devices and human users. We then identify a closed-form relationship between system reliability and wireless physical layer metrics  and derive a closed-form expression for the reliability as a function of the human brain's delay perception. Simulation results using real data show that the proposed brain-aware approach can substantially save power in the network while preserving the reliability of the users, particularly in low latency applications. In particular, the results show that the proposed brain-aware approach can yield power savings of up to $78$\% compared to a conventional, brain-unaware system.   
	
	The rest of the paper is organized as follows. Section
	\ref{sec:SysModel} introduces the system model. Sections III and IV present the proposed learning algorithm and resource allocation framework, respectively.
	Section V presents the simulation results and conclusions are drawn
	in Section VI.

	\section{System Model and Problem Formulation}\label{sec:SysModel}
	Consider the downlink of a  cellular network with humans-in-the-loop having a single base station (BS) serving a set $\mathcal{H}$ of $N$ human users with their UEs and a set $\mathcal{M}$ of  $M$ machine type devices (MTDs). {We assume that each human user uses one UE.} Each UE or MTD can have a different application with different QoS requirements such as sending a command to an actuator (for an MTD) or playing a 3D interactive game (for a UE).  We consider a time-slotted system, 
    and define $\mathcal{K}$ as the set of $K$ resource blocks (RBs).
	In our model, the packets associated with user  $i \in \mathcal{H}\cup \mathcal{M}$ arrive at the BS according to independent Poisson processes with rate $a_i(t)$. The lengths $l_i, \forall i \in \mathcal{H}\cup \mathcal{M}$ of the packets follow an exponential distribution.  Hence, each user's buffer at the BS will follow an M/M/1 queuing model. The total queuing and transmission delay of each user $i$ is $D_i(t)=q_i(t)+\frac{l_i}{r_i(t)}$, where $q_i(t)$ is the queuing delay.
	The data rate for each user is given by:
	\begin{equation}\label{eq:rateShannon}
	r_i(t)=B \sum_{j=1}^{K} \rho_{ij}(t)\,\log_2\left(1+\frac{p_{ij}(t) h_{ij}(t)}{\sigma^2}\right),
	\end{equation}
	where 	$p_{ij}(t)$ is the transmit power between the BS and user $i$ over RB $j$ at  time $t$ and $h_{ij}(t)$ is the time-varying Rayleigh fading channel gain. In (\ref{eq:rateShannon}), 
    $\rho_{ij}(t)=1$ if RB $j$ is allocated to user $i$ at time slot $t$, and $\rho_{ij}(t) =0$, otherwise.  $B$ is the bandwidth of each RB. {$\sigma^2$ is the noise power which is defined as the power spectral density of the noise multiplied by the bandwidth $B$.}

 We define $\beta_i(t)$ as the delay perception threshold for any user $i \in \mathcal{H}\cup \mathcal{M}$ at time $t$. If the delay decreases below the threshold $\beta_i(t)$, the user will not be able to discern the change in service quality.
We use the concept of delay perception $\beta_i(t)$ to measure time varying delay requirements of UEs and MTDs. Since the delay perception for MTDs is constant, hereinafter, for simplicity, we use $\beta_i(t)$ to exclusively denote the delay perception of human users, i.e, $\beta_i(t), \forall i \in \mathcal{H}$, unless mentioned otherwise.
 This delay perception can be affected by multiple sources pertaining to the human brain such as context, human attention, human fatigue or cognitive abilities and is determined by measuring the capabilities of the human brain at each time slot using machine learning methods.
	\emph{By explicitly accounting for the cognitive limitations of the human brain, the BS can better allocate resources to the users that need it, when they can actually use it}. This is in contrast to conventional brain-agnostic networks \cite{semiari2015context,huang2016system} in which resources may be wasted, as they are allocated only based on application QoS without being aware on whether the human user can indeed process the actual application's QoS target.

	We pose this resource allocation problem as a power minimization  problem that is subject to a brain-aware QoS constraint on the latency:
	\begin{subequations}
		\label{eq:OrigOptim}
		\begin{align}
		\min_{\rhob(t),\Pb(t)} \quad &\sum_{j \in \mathcal{K}}\Big[\sum_{i\in \mathcal{H}}  \bar P_i^j+\sum_{i\in \mathcal{M}} \bar P_i^j\Big], \label{eq:CostFcn}\\ 
		\text{s.t.} \qquad &\text{Pr}\big\{D_i(t)\geq D_i^{\max}\big(\beta_i(t)\big)\} \leq \epsilon_i\big(\beta_i(t)\big), &\forall i \in \mathcal{H} \cup \mathcal{M}, \label{eq:constrain1} \\
		&p_{ij}(t) \geq 0, \,\, \rho_{ij}(t) \in \{0,1\}\quad  &\forall i \in \mathcal{H} \cup \mathcal{M}, j\in \mathcal{K}\label{eq:constrain3},\\
        &\sum_{i \in \mathcal{H} \cup \mathcal{M}}\rho_{ij}(t)=1, \quad&\forall j\in \mathcal{K}\label{eq:constrain5},
		\end{align}
	\end{subequations}
	where $\rhob(t)$ is an $(M+N)\times K$ matrix having each element $\rho_{ij}(t)$.
	$\Pb(t)$ is an $(M+N)\times K$ matrix with each element $p_{i,j}(t)$ representing the instantaneous power allocated to user $i$ on RB $j$. The term $\bar{P}_i^j=\lim_{t\to \infty} \frac{1}{t} \sum_{\tau=0}^{t-1}\rho_{ij}(\tau) p_{ij}(\tau) $ is the time average of the power allocated to user $i$ on RB $j$.  
$D_i(t)$ incorporates both transmission and queuing delays.	$D_i^{\max}\big(\beta_i(t)\big)$ is the maximum tolerable delay. {This delay depends on $\beta_i(t)$ because changes in human delay perception will change maximum tolerable delay for human users.}  { $\epsilon_i(\beta_i(t))$ in equation (2b) denotes the maximum probability of the packet delay exceeding $D_i^{\max}(\beta_i(t))$. Hence, we can define $1-\epsilon_i(\beta_i(t))$ as the reliability of the user $i$}. We define \emph{reliability} as the proportion of time during which the delay of a given user does not exceed a threshold. {For notational convenience, hereinafter, we  use the terms  $D_i^{\max}\left(\beta_i(t)\right)$  and $D_i^{\max}$ interchangeably.}
{Constraint  (\ref{eq:constrain1})   takes into account the packet size and the rate of the application implicitly in addition to the maximum tolerable delay and reliability. From a resource allocation perspective, we can consider any application using  (\ref{eq:constrain1}).}

	The key difference between our problem formulation and conventional RB allocation problems \cite{Rui2017APCC} is  seen in the QoS delay requirement in (\ref{eq:constrain1}). 
Constraint (\ref{eq:constrain1}) is with respect to two random processes $D_i(t)$ and $\beta_i(t)$. In (\ref{eq:constrain1}), the network explicitly accounts for the human brain's (and the MTDs') delay needs. By taking into account the features of the brain of the human UEs, the network can avoid wasting resources. This waste of resources can stem from allocating more power to a UE, solely based on the application  QoS, while ignoring how the brain of the human carrying the UE perceives this QoS. Clearly, ignoring this human perception can lead to inefficient resource management.
	
\begin{figure}[!t]
    \begin{minipage}{0.49\textwidth}
	\centering
	\includegraphics[scale=.67,trim={0 0 0 0}]{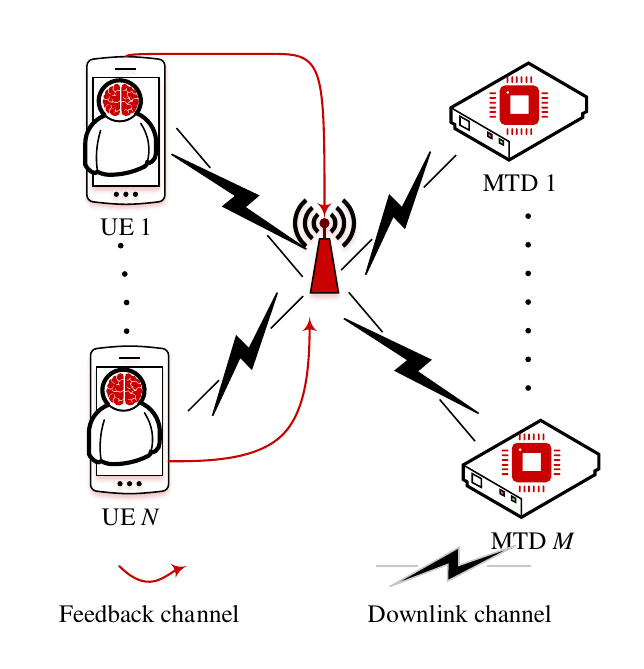}
	\caption{Illustration of the system model.}
	\label{fig:SysModel}

	\end{minipage}
    \hfill
\begin{minipage}{0.49\textwidth}
	\centering
	\includegraphics[scale=.75,trim={0 0 0 0}]{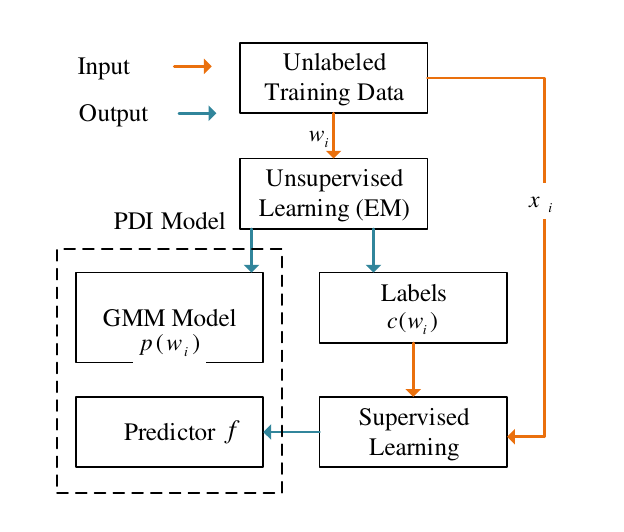}
	\caption{{Graphical representation of building a  PDI model.}}
	\label{fig:PDImodel}
        \end{minipage}
	\end{figure}

	{We propose a machine learning algorithm to identify the human brain delay perception $\beta_i(t)$.}
	Each human user has $d$ features, (e.g., age, occupation, location) assumed to be known to the BS. This time-varying feature vector is denoted by $\xb_i(t) \in \mathds{R}^d$. We develop a learning algorithm to build a model that maps these features to $\beta_i(t)$ for each user. We  then show that being aware of $\beta_i(t)$ can help the resource allocation algorithm to save a significant amount of resources for low-latency systems. 
	{We assume that the BS has access to the user features $\xb_i(t)$. In practice, the BS can collect such data whenever a given user registers in the network or by using the sensors of a user's mobile device. The system model is shown in Fig. \ref{fig:SysModel}}. Also, Table \ref{TablePara} provides a list of our main parameters and notations.

	{To find the mapping $\beta_i(t)=f(\xb_i(t))$ between human features $\xb_i(t)$ and the delay perception of the brain, we introduce a novel supervised learning mechanism called the \emph{probability distribution identification (PDI) method}. Here, function $f(.)$ shows this mapping. Since reliability is a key factor in a communication system, we need a supervised learning algorithm that not only predicts $\beta_i(t)$ as a function of $\xb_i(t)$, but also gives a measure of reliability for this prediction. This \emph{measure of reliability} is one of the key advantages of PDI learning over other supervised learning methods \cite{bishop2007pattern, chen2017machine}.} Although conventional methods such as neural networks can be used to approximate the continuous function $f(.)$ \cite{chen2017machine}, these methods cannot quantify the reliability of this prediction.  The reliability of predictions is defined as the probability that the prediction of $\beta_i(t)$ lies within a certain range of the true values for $\beta_i(t)$.
	
		 \begin{table}[!t]
 	\normalsize
 	\begin{center}{
 		\centering
 		\caption{ List of notations.}
 		\label{TablePara}
 		\resizebox{15cm}{!}{
 		{\hspace*{-1cm}

 			{\begin{tabular}{|c|c|c|c|}
 				\hline
 				\textbf{Notation} & \textbf{Description} &\textbf{Notation} & \textbf{Description} \\ \hline \hline
 				$p_{ij}$	&   Power of user $i$ on RB $j$ &$\rho_{ij}$ & RB $j$ allocation indicator   to user $i$         \\ \hline 
 				$D_i$	&    Packet delay for user $i$      & $L$ & Total number of clusters, brain modes \\ \hline
 				
 				$D_i^{\max}(.)$	&     Target delay for user &$\epsilon_i(.)$ & target reliability of user $i$   \\ \hline
 				
 				$\beta_i(t)$	&    Delay perception of user $i$ at time $t$   &$V$ &Balancing parameter for reliability constraint \\ \hline

 				$h_{ij}(t)$	&   Time-varying Rayleigh fading channel gain  &$\lambdab$ & Lagrange multiplier      \\ \hline
 				
 				$B$	&    RB bandwidth      &$\sigma^2$ & Noise power \\ \hline
 				
 				$N$	&     Number of UEs    &$k(j)$ & Optimal allocation for RB $j$\\ \hline
 				
 				$K$	&     Number of available RBs & $\Pi$ & Projection operator\\ \hline
 				
 				$a_i(t)$	& Arrival rate of user $i$ in time $t$   &$\zeta_i$ &Subgradient of element $i$ of $\lambda_b$   \\ \hline
 				
 				$F_i(t)$	&    Virtual queue for user $i$ at time $t$  &$\Sigmab_k$ & Covariance matrix of mode $k$ of human brain \\ \hline

 				$n$	&     \# of training users, &$\psi(.)$ & PDF of multivariate normal     \\ \hline

 				$D_i^{\max}$	&      Target end-to-end latency for user $i$  &$\mub_k$ & Mean vector    of mode $k$ of human brain\\ \hline
 				
 				$\mathcal{M}$	&      set of MTDs        &$\mathcal{H}$ &Set of UEs\\ \hline
 				$\mathcal{K}$	&      Set of RBs        &$\chi_{d+1}^2(.)$ &chi-square distribution function\\ \hline
 				
 					$\mathcal{\pi}_k$	&  Mixture weight of mode $k$        &$\zb$ &Mode indicator vector\\ \hline
 				$\mathcal{L}$	&      Likelihood function &$d$ &\# of features for each user\\ \hline
 				$f$	& Supervised learning model &$c(\wb_i),c_i$ & Cluster (mode) number of $\wb_i$ \\ \hline
 				$\yb$	& Set of labels for supervised learning &$\xi(.)$ & 0--1 loss function\\ \hline
 				
 				$D_i^{\min}(\epsilon')$	& Effective delay of human user $i$ &$Q_d(\gamma)$ & Quantile function of chi-square distribution with $d$ DOF\\ \hline
 				$\mathcal{S}$ & Reliability event of the system
 				&$\chi_d(.)$ &Chi-square function with $d$ DOF\\\hline
 				$\mathcal{L}_a$ & Lagrange dual function & $e_j$ & Unit vector at $j$\\\hline
 				
 		\end{tabular}}}
 		}
 	}
 	\end{center}
 \end{table}

	{The PDI method can find the distribution of the prediction values. Although many existing supervised learning methods can build a model for predicting an output based on a given input, they fail to find a statistical model for this prediction. In addition, by using the statistical model for the predictions s resulting from our proposed PDI, the system designer can better  design the system based on desired reliability. }

	As discussed in \cite{petkoski2015effects}, the delay perception of a human brain typically follows a multi-modal distribution. As a result, we design the proposed PDI approach to capture such a model and find the different modes of a human brain. Then, using the distribution of the brain delay, the PDI approach can find the \emph{effective delay}  of the human brain. This effective delay determines relationship between $\beta_i(t)$ and $\xb_i(t)$ along with its reliability.

    \subsection{{Building a PDI model}}	
	 Consider a dataset $\{\xb_1(t),\cdots,\xb_n(t)\}$, where $\xb_i(t)\in \mathds{R}^d$ is one sample data vector. The elements of $\xb_i(t)$ are {user} features which can be both categorical (such as gender) and numerical (such as age). For each input vector $\xb_i(t)$, we have a corresponding output value of delay perception $\beta_i(t)$.  This data can be collected using experiments or surveys such as those in \cite{yang2015prospect}. Since we can remove time dependency of the data using time-series techniques {such as in \cite{Baydogan2015}}, hereinafter, we use $\xb$ instead of $\xb(t)$. {Although we omit the time dependency from $\xb_i(t)$ for the training process, it is still implicitly a function of time.} This dataset can be represented by a matrix $\Xb \in \mathds{R}^{n\times d}$, where $\xb_i^T$ is row $i$ of $\Xb$.
	Using PDI, we first create an $n\times(d+1)$ dataset matrix:
	\begin{equation}\label{eq:define w}
	\boldsymbol{W}=[\boldsymbol{X} \Vert \boldsymbol{\beta}]=
	\begin{bmatrix}
	\boldsymbol{w}_1^T\\
	\vdots\\
	\boldsymbol{w}_n^T
	\end{bmatrix}=
	\begin{bmatrix}
	\xb_1^T &\beta_1(t)\\
	\vdots &\vdots\\
	\xb_n^T &\beta_n(t)\\
	\end{bmatrix},
	\end{equation}
	where $\boldsymbol w_i \in \mathds{R}^{d+1}$ is   a vector of the delay perception $\beta_i(t)$ and $d$ other correlated features of the human brain.
	{First, in the unsupervised learning step, we fit a Gaussian mixture model (GMM) to our dataset using the expectation-–maximization (EM) algorithm  \cite{moon1996expectation} to obtain $p(\xb_i,\beta_i(t))$.}
	{After finding $p(\xb_i,\beta_i(t))$, we are able to cluster the data samples and find $m$ brain modes in the data. Then, each data vector $\xb_i$ is labeled based on its cluster number $c_i$ so that each $\xb_i, \,i=1,\cdots,n$ has a label in the cluster set $c_i\in \mathcal{C}=\{1,\cdots,L\}$. Using this method we have a labeled dataset which can be used for the supervised learning.
	\emph{These cluster numbers  will correspond to the modes of the human brain that determine its effective delay perception.} 
	}

	{Next, we describe the Gaussian mixture model that we use as underlying model for the human brain. We use GMM for clustering (unsupervised learning) and statistical modeling of human brain because of its multimodal structure which resembles human brain activites \cite{jaimes2007multimodal},} its scalibility, and its robustness and stability under high-noise levels compared to nonparametric methods.

	A multi-modal stochastic model is assumed for the brain features $\wb_i$ for user $i$. The proposed distribution for $\boldsymbol w_i$ is given by \cite{bishop2007pattern}:
	\begin{equation}
	\label{eq:multi modal distribution}
	p(\boldsymbol w_i)=\sum_{z} p(\zb)p(\boldsymbol w_i|\zb)=\sum_{k=1}^L \pi_k \psi(\boldsymbol w_i|\boldsymbol{\mu}_k,\boldsymbol\Sigmab_k),
	\end{equation}
	where $\psi(\boldsymbol w_i|\mub_k,\Sigmab_k)$ is the probability density function for a multivariate normal distribution with mean vector $\mub_k$ and covariance matrix $\Sigmab_k$. 
	$\Sigmab_k$ and $\boldsymbol{\mu}_k$ represent the covariance matrix and mean vector for mode $k$ of the human brain, respectively. {$\boldsymbol{z}$  is a binary random vector, in which a particular element $z_k$ is equal to $1$ and all other elements are 0. $\boldsymbol{z}$ essentially indicates which mode is activated in the GMM, and $\pi_k$ is defined as 
	$p(z_k=1)=\pi_k$. 
	$L$ is the total number of modes in the GMM.} The human brain will be in mode $k$ with probability $\pi_k$, and its features are generated using a multivariate normal distribution with mean and covariance $\boldsymbol{\mu}_k$ and $\Sigmab_k$, respectively. 
	The posterior probability, i.e. \emph{responsibility},  for mode $k$ will be:
	{
	\begin{equation}\label{eq:responsblty}
	r_i(z_k)=\frac{\pi_k \psi(\boldsymbol{w}_i|\boldsymbol{\mu}_k,\Sigmab_k)}{\sum_{j=1}^L {\pi_j} \psi(\boldsymbol{w_i}|\boldsymbol{\mu}_j,\Sigmab_j)}.
	\end{equation}}
	This responsibility can be used for clustering the data as well. After fitting the GMM on the dataset, we can find the mode with highest responsibility for each data point and assign the data to this mode. {The EM algorithm is used to find $\mub_k$, $\Sigmab_k$, and $\pi_k$ for all $k=1,\cdots,L$, based on the real-time human brain behavior \cite{moon1996expectation}.} The log likelihood function for our dataset can be written as:
	{
	\begin{equation}\label{eq:loglikelihood}
	\ln \mathcal{L}(\Sigmab, \mu, \pi|\wb) =\ln \sum_i p(\wb_i| \Sigmab, \mu, \pi) =\sum_i \ln \sum_{k=1}^L \pi_k \psi(\boldsymbol w_i|\boldsymbol{\mu}_k,\boldsymbol\Sigmab_k). 
	\end{equation}
	}
	The likelihood function in (\ref{eq:loglikelihood}) has singularities and, hence, it is infeasible to find  parameters $\pi_k$, $\Sigmab_k$, and $\mub_k$. The EM algorithm is proposed in \cite{EM98main} to maximize the likelihood function for a Gaussian mixture model. { In the EM algorithm, we first initialize $\Sigmab_k$, $\boldsymbol{\mu}_k$, and $\pi_k$ randomly. Next, we find the responsibility for each mode using (\ref{eq:responsblty}). Then, we reestimate parameters using current responsibilities. Finally,  the likelihood in (\ref{eq:loglikelihood}) is maximized with respect to  $\Sigmab_k$, $\boldsymbol{\mu}_k$ ,and $\pi_k$.}
	
    {As a result of the EM algorithm (unsupervised learning), we now have a GMM of our dataset matrix $\boldsymbol{W}$. 
	Based on this GMM, the data will be labeled (clustered) as follows. 
	For each data point $\wb_i$, the most probable mode is assigned as the label of this data, i.e., \begin{equation}\label{eq:ClusterAssignment}
    c_i=c(\wb_i)=\underset{k}{\arg \max}\,\, p(z_k=1|\wb_i)=\underset{k}{\arg \max}\,\, r_i(z_k).
    \end{equation}  
	In (\ref{eq:ClusterAssignment}), we assign the most likely cluster to each data point $\wb_i$.}
	{After building a GMM model using unsupervised learning on the $\boldsymbol{W}$ dataset to obtain target vector $\yb=\begin{bmatrix}
	c(\wb_1) & \cdots &c(\wb_n)
	\end{bmatrix}^T$, we use the pair $\{\Xb,\yb\}$ to train a supervised learning model.
	Thus, the output of the unsupervised learning step $\yb$ is used for training the supervised learning model.
	Then, during the supervised learning step, we train a classifier so that it can find the mode $c_i$ using the human features $\xb_i$ as input.
    Given the data matrix $\Xb$ and the output vector $\yb$, this supervised learning builds us a model $f$ such that $c_i=f(\boldsymbol{x_i})$, where
    \begin{equation}\label{eq:supervisedF}
    f=\textrm{arg}\min_{\hat f} \sum_{i=1}^n \xi\left(c(w_i),\hat f (x_i)\right),
    \end{equation}
    where $\xi(.)$ is a 0-1 loss function \cite[Equation 7.5]{friedman2001elements}. $f$ is a function that is approximated using a set of points $(\xb_i,c_i)$ and determines the relationship between the features of a user and its cluster.} To overcome overfitting, we use the elbow method for finding the optimal number of clusters in PDI, as discussed in Section \ref{sec:Simul}.
	After approximating $f$, given each human user's feature vector $\xb_i$, we find the modes $c_i$ using model $f$.
	{Algorithm 1 summarizes building a PDI learning model.} {Also, a graphical representation for building a PDI model is shown in Fig. \ref{fig:PDImodel}}.

\subsection{{Deployment of the PDI learning model}}
\setlength{\textfloatsep}{0pt}%
\begin{algorithm}[!t]
 \caption{{Building PDI model}}
 {
 \begin{algorithmic}[1]
 \renewcommand{\algorithmicrequire}{\textbf{Input:}}
 \renewcommand{\algorithmicensure}{\textbf{Output:}}
 \REQUIRE $w_i=[x_i \quad\beta_i],\, i =1,\cdots,n$
 \ENSURE  $f,\, \pi_k,\, \mu_k,\, \Sigma_k, \quad k=1=\cdots,L$
 \\ {\textit{Unsupervised Learning}} :
  
  \STATE Apply EM algorithm to $w_i$, find $\pi_k,\, \mu_k,\, \Sigma_k,\, r_i(z_k)\quad  k=1,\cdots,L,\quad i=1 \cdots,n.$ 
  \FOR {$i=1=\cdots,n$} 
  \STATE find $c(\wb_i)$ using (\ref{eq:ClusterAssignment}).
  \ENDFOR
  {\STATE Pass $\yb=\begin{bmatrix}
	c(\wb_1) & \cdots &c(\wb_n)
	\end{bmatrix}^T$ to the supervised learning algorithm
  \\ \textit{Supervised Learning} :
  \STATE Find $f$ using (\ref{eq:supervisedF}).}
 \RETURN $f,\, \pi_k,\, \mu_k,\, \Sigma_k, \quad k=1=\cdots,L$
 \end{algorithmic} }

 \end{algorithm}

	{In this subsection}, we bound $D_i^{\max}(\beta_i(t))$ based on its features $\xb_i$. {Note that the deployment and training of the PDI learning method are separated from each other. The deployment part only uses the model generated by the training part.} Now that the system can identify the human users' modes, we need to find a relationship between a human user's mode and the probabilistic model of its delay perception by defining the concept of effective delay.

	\begin{definition}
		Given the statistical model for human  delay perception $\beta_i(t)$, $D_i^{\min}(\epsilon')$ is the \emph{effective delay} for human user $i$ that satisfies: 
		\begin{equation}\label{eq:dminKia}
		\text{Pr}\big\{\beta_i(t)<D_i^{\min}(\epsilon')\big\}<\epsilon'.
		\end{equation}
	\end{definition}

	To  find the effective delay for human user $i$, we first find the probability that the  delay perception of human user $i$ is less than a threshold $D_i^{\min}(\epsilon')$. In other words, we want to find the relation between $\epsilon'$ and $D_i^{\min}(\epsilon')$ in (\ref{eq:dminKia}).  The concept of effective delay is defined using the fact that delays less than $D_i^{\min}(\epsilon')$ cannot be sensed by a human with ($1-\epsilon'$) certainty. The relation between $\epsilon'$ and $D_i^{\min}(\epsilon')$ in (\ref{eq:dminKia}) is found in Theorem \ref{lemma:confidence region single mode}. {For notational simplicity, hereinafter, we use $D_i^{\min}$ instead of $D_i^{\min}(\epsilon')$.}

	\begin{theorem}\label{lemma:confidence region single mode}
If brain mode $k$  is identified for user $i$, then its delay perception will be bounded as follows:
	\begin{equation}\label{eq:lemma1 main result}
	\text{Pr}\Big\{|\beta_i(t)-\mu_k(d+1)|<\sqrt{Q_{d+1}(\gamma) \boldsymbol{e}_{d+1}^T\Sigmab_k \boldsymbol{e}_{d+1}}\Big\}>\gamma,
	\end{equation}
   { where $\Sigmab_k$ and $\mu_k(d+1)$ represent, respectively, the covariance matrix  and the $(d+1)$th element of the mean vector of  the identified brain mode $k$.}
		{$Q_d(\gamma)$ is the quantile function of chi-square distribution with $d$ degrees of freedom,} and is defined as
		\begin{equation}
		Q_{d+1}(\gamma)=\inf\Big\{x\in \mathds{R}| \gamma\leq \int_{0}^{x} \chi_{d+1}^2(u) du\Big\},
		\end{equation}
		and	$\boldsymbol{e}_j$ is a unit vector in $\mathds{R}^{d+1}$, whose $j$th element is $1$ and all other elements are zero. $d$ is number of features  used for learning. $\chi_{d+1}^2(x)$ is the probability density function of a chi-square random variable with $d+1$ degrees of freedom. 
	\end{theorem}
\begin{IEEEproof}
	See Appendix A.
\end{IEEEproof}

\begin{algorithm}[!t]
 \caption{{Deploying PDI model for finding brain delay perception}}
 {
 \begin{algorithmic}[1]
 \renewcommand{\algorithmicrequire}{\textbf{Input:}}
 \renewcommand{\algorithmicensure}{\textbf{Output:}}
 \REQUIRE $x_i$, $i=1,\cdots,N$
 \ENSURE  $\mu_i,\,\Sigma_i$, $i=1,\cdots,N$

  \FOR {$i=1=\cdots,N$} 
  \STATE find the cluster for $x_i$: $k_i=f(x_i)$ and hence $\mu_{k_i},\Sigma_{k_i}$.
  \STATE find function  $D_i^{\min}(\epsilon')$ using (\ref{eq:DminVepsilon})
  \STATE using target reliability find $\epsilon'$ and then using function  $D_i^{\min} (\epsilon')$ find $D_i^{\min}$.
  \STATE set the $D_i^{\max}(\beta_i(t))=D_i^{\min}$, and using  (\ref{eq:findAllthings}) and $\epsilon'$ find $\epsilon(\beta_i(t))$ 
  \ENDFOR
 \RETURN $D_i^{\max}(\beta_i(t))$ and $\epsilon(\beta_i(t))$ for  $i=1=\cdots,N$
 \end{algorithmic} }
 \end{algorithm}

\begin{figure}[!t]
    \begin{minipage}{0.49\textwidth}
	\centering
	\includegraphics[scale=.67,trim={0.55cm 0 0 0}]{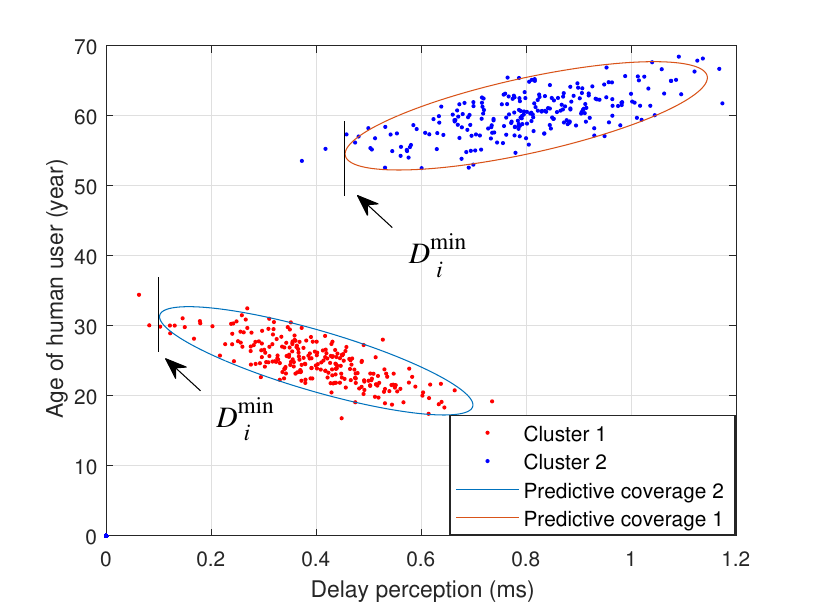}
	\caption{Finding $D_i^{\min}$ using a GMM model for two different clusters. }
	\label{fig:DuBa}

	\end{minipage}
    \hfill
\begin{minipage}{0.49\textwidth}
	\centering
	\includegraphics[scale=.5,trim={0.6cm 0 0 0}]{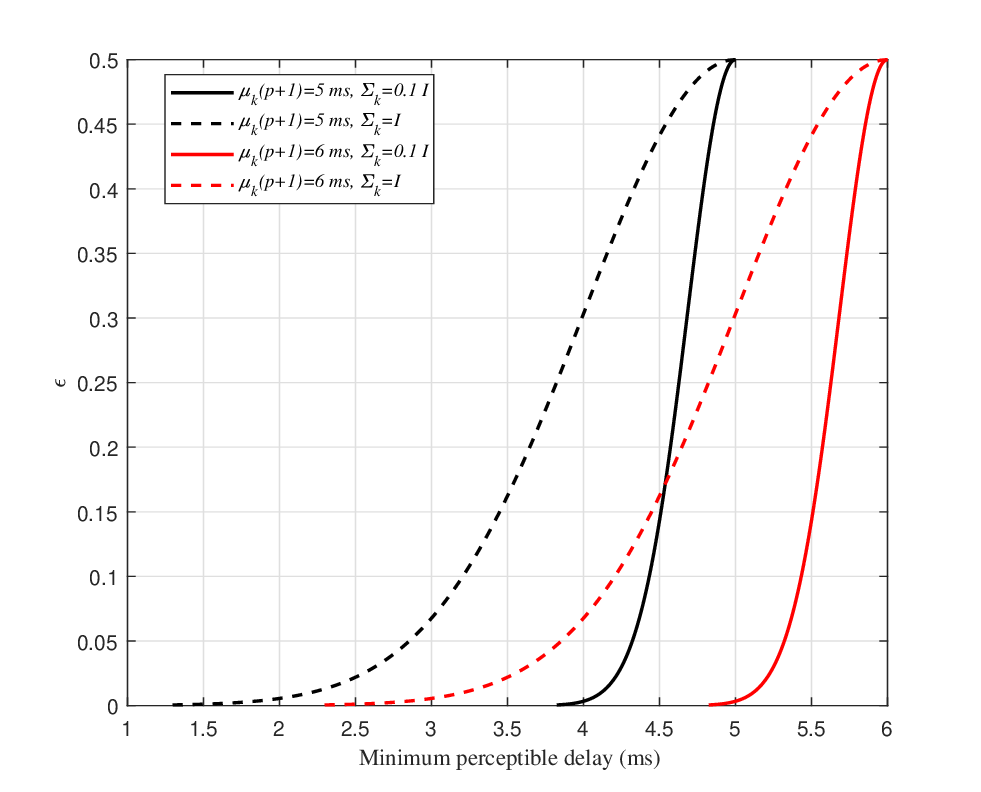}
	\caption{ٍRelationship between $\epsilon$ and $D_i^{\min}(\epsilon)$ for different values of $\mu_k(d+1)$ and $\Sigmab_k$. $I$ is the identity matrix.}
	\label{fig:Delay versus epsilon}

        \end{minipage}
	\end{figure}

Note that the bound in (\ref{eq:lemma1 main result}) is different from finding a bound  using marginal distributions, as it is based on high probability density areas. The use of a marginal distribution is not possible here, since the Gaussian assumption is only valid locally around the mean. Also, in case of data classification error which mostly happens when the data is located in the overlapping area between clusters, the bound in (\ref{eq:lemma1 main result}) will be either more conservative than the actual bound for delay perception or it will not change significantly compared to the actual bound. As seen from Theorem \ref{lemma:confidence region single mode}, in addition to the delay perception element  $\mu_k(d+1)$, the only other parameter that affects the delay is $\boldsymbol{e}_{d+1}^T\Sigmab_k \boldsymbol{e}_{d+1}$, which is the $(d+1)$th diagonal element of $\Sigmab_k$, which is not assumed to be diagonal.
Fig.\ref{fig:DuBa} shows the relationship between $D_i^{min}$ and {GMM} random data generated from a Gaussian mixture model.  Fig.\ref{fig:DuBa} shows that, after finding the GMM for the dataset, one can find the predictive coverage of each Gaussian distribution and, then, we can determine the probability with which $\beta_i(t)$ for a user $i$ will be higher than a threshold $D_i^{\min}$. 
In order to find the effective delay for human user $i$, we first find the probability with which the  delay perception for human user $i$ will be less than a threshold $D_i^{\min}$. In other words, we will find the relationship between $\epsilon$ and $D_i^{\min}$ in (\ref{eq:dminKia}) using the following corollary that follows directly from Theorem 1.

\begin{corollary}
	As a direct result of Theorem \ref{lemma:confidence region single mode}, we can reduce  (\ref{eq:lemma1 main result}) to
	\begin{equation}\label{eq:Dmin epsilon}
	\text{Pr}\Big\{\beta_i(t)<\mu_k(d+1)-\sqrt{Q_{d+1}(\gamma) \boldsymbol{e}_{d+1}^T\Sigmab_k \boldsymbol{e}_{d+1}}\Big\}<\frac{1-\gamma}{2}.
	\end{equation}
	Therefore, we find $D_i^{\min}(\epsilon)$ and $\epsilon$ in (\ref{eq:dminKia}) as 
	\begin{equation}\label{eq:DminVepsilon}
	D_i^{\min}(\epsilon)=\mu_k(d+1)-\sqrt{Q_{d+1}(1-2\epsilon) \boldsymbol{e}_{d+1}^T\Sigmab_k \boldsymbol{e}_{d+1}}.
	\end{equation}
\end{corollary}

Since $Q_{d+1}(\gamma)$ can only be calculated numerically, a closed-form relationship between $D_i^{\min}(\epsilon)$ and $\epsilon$ cannot be found. However, we can numerically analyze this relationship, as shown in Fig. \ref{fig:Delay versus epsilon}.
{Fig. \ref{fig:Delay versus epsilon} is found using a set of points generated with (\ref{eq:DminVepsilon}) for different values of $\mu_k(d+1)$ and $\Sigmab_k$. } From Fig. \ref{fig:Delay versus epsilon}, we can first observe that $D_i^{\min}(\epsilon)$ is an increasing function. \emph{This means that the probability of the human brain noticing QoS differences for low delays will be much smaller than for higher delays}, which is an intuitive fact. Furthermore, it can be inferred that, if the delay perception for a group of human users within a cluster is diverse, then the system's confidence on the delay perception of this group of humans will decrease, i.e., the estimation of the delay perception of this group of human users will be less reliable. Next, we determine constraint (\ref{eq:constrain1}) using  $D_i^{\min}(\epsilon)$.

As stated before, some delays are not perceptible to human users. To capture this feature, we find $D_i^{\max}(\beta_i(t))$ and $\epsilon(\beta_i(t))$ in problem (\ref{eq:OrigOptim}) using $D_i^{\min}(\epsilon)$. Recall that $D_i^{\max}(\beta_i(t))$ is a parameter that will be used by the resource allocation system to represent the maximum tolerable delay for the reliable communication of user $i$ with $1- \epsilon(\beta_i(t))$ being the reliability of user $i$.
There are three possible cases for $D_i^{\max}$ based on $D_i^{\min}$ of a human user $i$:
\subsubsection{$D_i^{\max}>D_i^{\min}$}   In this case, the system will not be reliable even if we satisfy $\text{Pr}(D>D_i^{\max})<\epsilon$. The reason is that the human user has a delay perception of less than the maximum delay $D_i^{\max}$ and hence, the system is not reliable.

\subsubsection{$D_i^{\max}<D_i^{\min}$} In this case, if the system is able to satisfy $\text{Pr}(D>D_i^{\max})<\epsilon$, then the system will be reliable, because user $i$ cannot sense delays less than $D_i^{\min}$ and its service delay will not exceed $D_i^{\max}$.

\subsubsection{$D_i^{\max}=D_i^{\min}$} If this equality holds, the system will be reliable and  it will also have prevented a waste of resources. If any given user cannot perceive delays less than $D_i^{\min}$, then it is not effective to allocate more resources to this user.

{We define $\mathcal{S}$ as the event resulting from case 2 and case 3}
{while assuming events $E_1$ and $E_2$ satisfy $D<D_i^{\max}$ and $\beta_i(t)>D_i^{\min}$, respectively.} We know that for case 1,  event $E_1\cap E_2 $ is a subset of  event $\mathcal{S}$, and in case 2,  event $\mathcal{S}$ is a subset of  event $E_1\cap E_2 $. Similarly, in  case 3,  event $E_1\cap E_2 $ is  same as event $\mathcal{S}$. Since the probability of $E_1\cap E_2 $ can be computed, if we set $D_i^{\min}$ to $D_i^{\max}$ (case 3), {we can find $\mathcal{S}$ as follows:}
\begin{align}
&\text{Pr}(E_1\cap E_2 )=1-\text{Pr}\Big((D>D_i^{\max}) \cup (\beta_i(t)<D_i^{\min})\Big)\label{eq:De Morgan}\\
&=1-\Big(\text{Pr}(D>D_i^{\max})+\text{Pr}(\beta_i(t)<D_i^{\min})-\text{Pr}(D>D_i^{\max}) \text{Pr}(\beta_i(t)<D_i^{\min})\Big). \label{eq:independent event}
\end{align}

(\ref{eq:De Morgan}) follows from De Morgan's law, and (\ref{eq:independent event}) is true since $D$ and $\beta_i(t)$ are two independent random variables {at each iteration}.
Therefore, if $D_i^{\min}=D_i^{\max}$ for user $i$ and $\epsilon\epsilon'$ is small,  we can see that
\begin{equation}\label{eq:findAllthings}
1-\Big(\text{Pr}(D>D_i^{\max})+\text{Pr}(\beta_i(t)<D_i^{\min})\Big)\geq 
1- (\epsilon +\epsilon'),
\end{equation}
and, hence, {
$
\text{Pr}(\mathcal{S})=\text{Pr}(E_1 \cap E_2)> 1- (\epsilon +\epsilon'),
$}
where {Pr}($\mathcal{S}$) is the reliability of the system defined in (\ref{eq:OrigOptim}).
Subsequently, as we design the system, we consider the reliability as a predetermined target design parameter for the system. Using this parameter, we can set $\epsilon$ and $\epsilon'$. Given $\epsilon'$ and numerical function $D_i^{\min}(\epsilon')$  derived in (\ref{eq:Dmin epsilon}), $D_i^{\min}$ can be determined. 
Now, given $\epsilon(\beta_i(t))$ and $D_i^{\max}(\beta_i(t))$, we can fully characterize problem (\ref{eq:OrigOptim}).
We note that, as the human user delay perception changes throughout a day, $D_i^{\max}(\beta_i(t))$ changes accordingly. Therefore, our solution should take into account changes in $D_i^{\max}(\beta_i(t))$ as well as changes in channel gains $h_{ij}(t)$. { $D_i^{\max}$ is a threshold and $1- \epsilon$ denotes the reliability which is used in all cellular systems.  However, in a wireless system that explicitly takes into account human users in its loop, $D_i^{\max}$ will become a function of $\beta_i(t)$ and can be determined using the concept of effective delay  $D_i^{\min} (\epsilon')$ defined in (9).}

    \section{Brain-Aware Resource Management}\label{sec:BrainAwRA}
	{The notion of human-in-the-loop implies that human factors (such as brain limitations) will be part of the resource allocation framework, i.e., in the loop of resource allocation. For such a system, resource management can dynamically adapt to the human user in its loop, as opposed to just the device. Therefore, our approach considers the human brain limitations as functions of time and adapts the system to the dynamic changes that can occur in the brain and its cognitive limitations, over time.}
	
	{Since the human brain state usually changes rapidly, our work is different from context-aware or QoS-aware works. The fast fluctuations in the cognitive activities of a human brain which have been validated in many works such as \cite{HARI2000455,fox2007spontaneous,laufs2003electroencephalographic} requires the resource allocation framework to be aware of time-varying brain-aware delay constraint in each time slot.}
	
	To solve problem (\ref{eq:OrigOptim}), we propose a novel brain-aware resource management framework that takes into account the time-varying wireless channel and the time-varying brain-aware delay constraint (\ref{eq:constrain1}). 
	{We transform this constraint into a mathematically tractable form.}

	 {The relation between the packet
 length distribution and the service time distribution for a packet is shown next, in Corollary \ref{corol:exponential service time}.} { Here, the packet service time is defined as the transmission time of a packet from the BS to the UE or MTD.}
	\begin{corollary}\label{corol:exponential service time}
		If a fixed rate $r_i$ is allocated to a user and the packet lengths follow an exponential distribution with parameter $\chi$, then, the distribution of the service time $s$ will also be exponential with parameter $\chi r_i$.
	\end{corollary}
	\begin{IEEEproof}
		{The CDF of the exponential distribution is $\mathcal{F}_{l}(\psi)=\text{Pr}(l<\psi)=1-e^{-\chi \psi}$. Hence, 
		\begin{equation}
		\mathcal{F}_s(S)=\text{Pr}(s<S)=\text{Pr}(\frac{l}{r_i}<S)=\text{Pr}(l<r_i S)=\mathcal{F}_{r_i S}(s)=1-e^{-\chi r_i S}.
		\end{equation}}
		This means that the PDF for the service time is $f_S(s)=e^{-\chi r_i s}$.
	\end{IEEEproof}
	{the packet length distribution parameter $\chi$ is constant in our analysis,} without loss of generality, we assume that the service time of each packet is an exponential random variable with parameter $r_i$, which is the same as the rate allocated to the user. 
	We assume that for any given user, the packets arrive {according to a Poisson process } with the rate $a_i(\tau)$, and the user data rate is {exponential with parameter} $r_i(\tau)$ in  slot  $\tau=1,\cdots,t$. Next, we derive the probability with which the delay of a given user $i$ exceeds a threshold $D_i^{\max}$.
	\begin{theorem} \label{prop:packet length}
		Assume that user $i$ has a time varying rate $r_i^{\tau}$ at time slot $\tau$. If the duration  of each time slot is long enough for the queue to reach its steady state, i.e., 
		\begin{equation}\label{eq:time slot condition}
		\frac{1}{r_i(\tau)-a_i(\tau)}<<\delta \tau,
		\end{equation}
		then, the probability that the delay exceeds a threshold is 
		\begin{equation}\label{eq:time average delay}
		\text{Pr}(D>D_i^{\max})=	\lim_{t\to \infty} \frac{1}{t} \sum_{\tau=1}^{t} e^{-\big(r_i(\tau)-a_i(\tau)\big)D_i^{\max}},	
		\end{equation}
		under the condition that $r_i(\tau)>a_i(\tau)$ for all $\tau>0$.
	\end{theorem}
	\begin{IEEEproof}
		See Appendix B.
	\end{IEEEproof}

	Theorem \ref{prop:packet length} shows that constraint (\ref{eq:constrain1}) is satisfied if the network can satisfy the following condition 
	\begin{equation}\label{eq:2b transformed}
	\lim_{t\to \infty} \frac{1}{t} \sum_{\tau=1}^{t} e^{-\big(r_i(\tau)-a_i(\tau)\big)D_i^{\max}}<\epsilon.
	\end{equation}
	
\begin{figure}[!t]
		\centering
		\includegraphics[scale=.6,trim={0.6cm 0 0 0}]{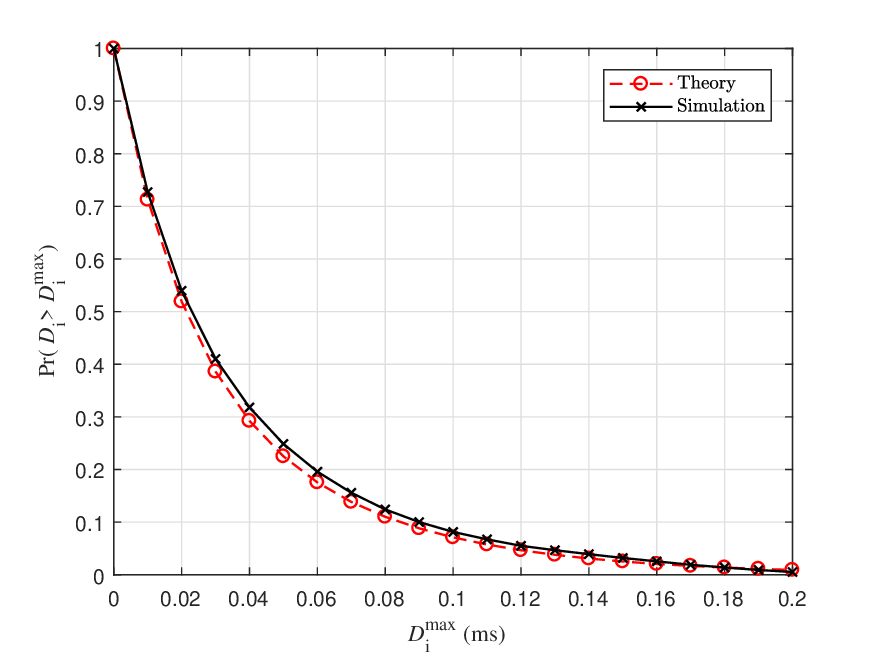}
		\caption{ٍComparison between simulation results and the result of Theorem \ref{prop:packet length}. }
		\label{fig:proposition}
	\end{figure}

    Fig. \ref{fig:proposition} shows the relationship between the theoretical result from Theorem \ref{prop:packet length} and simulation results. Clearly, simulation and analytical results are a near-perfect match with a maximum error of only $0.0146$.
    
	\subsection{Optimal Resource Allocation with Guaranteed Reliability}
Constraint (\ref{eq:2b transformed}) is analogous to the drift-plus-penalty method in  Lyapunov optimization framework \cite{neely2010stochastic} which we use to solve (\ref{eq:OrigOptim}). The problem has a time-varying nature since the human brain conditions and needs will change from time to time. The users' processing state $\beta(t)$ is also a function of time, and accordingly, the latency needs in (\ref{eq:constrain1}) will be time-varying. Therefore, we need to solve the optimization problem (\ref{eq:OrigOptim}) during each time slot efficiently. We propose an algorithm with a  low  computational complexity for solving problem (\ref{eq:OrigOptim}).
The drift-plus-penalty approach is used to stabilize a queue network while minimizing time average of a penalty function. To satisfy constraint (\ref{eq:constrain1}) in all time slots, (\ref{eq:time average delay}) must be smaller than $\epsilon$. {For this reason, we use virtual queues to model the time average constraint (\ref{eq:2b transformed}) in the optimization problem.} We define a virtual queue:
\begin{equation}\label{eq:F_iEq}
F_i(t+1)=\max\{F_i(t)+e^{-\big(r_i(t)-a_i(t)\big)D_i^{\max}}-\epsilon,0\}.
\end{equation}
 We can see that $e^{-\big(r_i(t)-a_i(t)\big)D_i^{\max}}-\epsilon<F_i(t)-F_i(t)$.
Consequently, we obtain:
{
\begin{equation}
\sum_{\tau=1}^{t}e^{-\big(r_i(\tau)-a_i(\tau)\big)D_i^{\max}}-\epsilon t<F_i(t)-F_i(0).
\end{equation}}
If $F_i(0)$ is bounded, we have:
{\begin{equation}
\lim_{t\to \infty} \frac{1}{t}\sum_{\tau=1}^{t}e^{-\big(r_i(\tau)-a_i(\tau)\big)D_i^{\max}}-\epsilon<\lim_{t\to \infty}\frac{F_i(t)}{t}.
\end{equation}}
If the queue $F_i(t)$ is mean-rate stable, that is, $\lim_{t\to \infty}\frac{F_i(t)}{t}=0$, then we have:
{\begin{equation}
\lim_{t\to \infty} \frac{1}{t}\sum_{\tau=1}^{t}e^{-\big(r_i(\tau)-a_i(\tau)\big)D_i^{\max}}<\epsilon.
\end{equation}}
The Lyapunov function is defined for all the queues in the base station as 
$
Y(t)=\frac{1}{2}\sum_{i\in \mathcal{M} \cup H} F_i(t)^2.
$
Then, we can find the drift function $\Delta t=Y(t+1)-Y(t)$ as: 
{\begin{equation}
Y(t+1)=\frac{1}{2}\sum_{i\in \mathcal{M} \cup \mathcal{H}} F_i(t+1)^2 \leq \frac{1}{2}\sum_{i\in \mathcal{M} \cup \mathcal{H}}F_i(t)^2+\frac{1}{2}\sum_{i\in \mathcal{M} \cup \mathcal{H}}y_i(t)^2+\sum_{i\in \mathcal{M} \cup \mathcal{H}}y_i(t)F_i(t),
\end{equation}}
where {\begin{equation}\label{eq:definition yt}
y^c_i(t)=e^{-\big(r_i(t)-a_i(t)\big)D_i^{\max}}-\epsilon.
\end{equation}}
Thus,
{\begin{equation} \label{eq:lyapunov opt ineq}
\Delta t \leq \frac{1}{2}\sum_{i\in \mathcal{M} \cup \mathcal{H}}y^c_i(t)^2+\sum_{i\in \mathcal{M} \cup \mathcal{H}}y^c_i(t)F_i(t).
\end{equation}}

We can form the drift-plus-penalty by adding $V\sum_{i,j}p_{ij}(t)$ to both sides of  inequality (\ref{eq:lyapunov opt ineq}), where $\sum_{i,j}p_{ij}$ is the total power of the BS which we want to minimize, and $V$ is a parameter that determines how important minimizing the objective function (\ref{eq:CostFcn}) is in comparison with satisfying (\ref{eq:constrain1}). We can balance the tradeoff between power and delay. The drift-plus-penalty inequality is 
{\begin{equation}\label{eq:before drift plus pealty}
\Delta t +V\sum_{i,j}p_{ij}\leq \frac{1}{2}\sum_{i\in \mathcal{M} \cup \mathcal{H}}y^c_i(t)^2+V\sum_{i,j}p_{ij}(t)+\sum_{i\in \mathcal{M} \cup \mathcal{H}}y^c_i(t)F_i(t).
\end{equation}}
Given that we assumed $r_i(t)>a_i(t)$ for all $t$, we know that {
$
|y^c_i(t)|<1 \, \forall t, i \in \mathcal{M} \cup \mathcal{H},
$}
and hence, we can rewrite (\ref{eq:before drift plus pealty}) as
{\begin{equation}\label{eq:drift plus pealty}
\Delta t +V\sum_{i,j}p_{ij}\leq UB+V\sum_{i,j}p_{ij}(t)+\sum_{i\in \mathcal{M} \cup \mathcal{H}}y^c_i(t)F_i(t),
\end{equation}}
where {$UB$} is the upper bound of {$\frac{1}{2}\sum_{i\in \mathcal{M} \cup \mathcal{H}}y^c_i(t)^2$}, and is equal to $\frac{|\mathcal{H}|+|\mathcal{M}|}{2}$. $|\mathcal{H}|$ is the cardinality of  set $\mathcal{H}$. Using the drift-plus-penalty algorithm \cite{neely2008fairness}, we know that, by minimizing the right hand side of equation (\ref{eq:drift plus pealty}),   queue $F_i(t)$ will be mean-rate stable, and hence, the condition { $y^c_i(t)<0$} will be satisfied. As a result, constraint (\ref{eq:constrain1}) will also be  satisfied. Furthermore, we know that by minimizing the right hand side of (\ref{eq:drift plus pealty}), cost function (\ref{eq:CostFcn}) is also minimized, owing to the fact that (\ref{eq:CostFcn}) is defined as a penalty function.
By minimizing the right hand side of (\ref{eq:drift plus pealty}), our optimization problem can be converted to the following time-varying problem:
{
\begin{subequations}\label{eq:converted optimization problem}
	\begin{align}
	\min_{\rhob(t),\Pb(t)} \qquad &V\sum_{i,j}p_{ij}(t)+\sum_{i\in \mathcal{M} \cup \mathcal{H}}y^c_i(t)F_i(t), \label{eq:converted cost func}\\
	\text{s.t.} \qquad &r_i(t)>a_i(t), \quad  &\forall i \in \mathcal{H} \cup \mathcal{M} \label{eq:converted constrain 1}\\
	&p_{ij}(t) \geq 0,\quad \rho_{ij}(t) \in \{0,1\}, \quad  &\forall i \in \mathcal{H} \cup \mathcal{M}, j\in \mathcal{K},\label{eq:converted constrain 2}\\
     &\sum_{i \in \mathcal{H} \cup \mathcal{M}}\rho_{ij}(t)=1, \quad&\forall j\in \mathcal{K}\label{eq:converted constrain 4}.
	\end{align}
\end{subequations}
}

The cost function in (\ref{eq:converted cost func}) is equivalent to (\ref{eq:CostFcn}) and (\ref{eq:constrain1}) in the original optimization problem. Learning the effective delay of each human user using our proposed PDI method determines the parameters {$y^c_i(t)$} and $F_i(t)$ in the problem (\ref{eq:converted cost func}). However, in order to satisfy (\ref{eq:constrain1}), we need to also satisfy (\ref{eq:converted constrain 1}). The reason for adding (\ref{eq:converted constrain 1}) is that if this constraint is not satisfied in any time slot, the queue length will approach infinity. Constraints (\ref{eq:converted constrain 2}) and (\ref{eq:converted constrain 4}) are feasibility conditions and remain the same as (\ref{eq:OrigOptim}).
Hence, by solving (\ref{eq:converted optimization problem}) in each time slot, the original problem (\ref{eq:OrigOptim}) will be solved. 

Nonetheless,  
 problem (\ref{eq:CostFcn})  is not a convex optimization problem, due to the fact that it is a mixed integer problem  and its complexity increases exponentially with the number of users. Since (\ref{eq:CostFcn}) needs to be solved at each time slot, this exponential order of complexity makes the implementation infeasible. Consequently, we should use a dual decomposition method to break down optimization problem (\ref{eq:converted optimization problem}) to smaller subproblems, and find the optimal solution to (\ref{eq:converted optimization problem})  using a low complexity method.
It is rather challenging to solve (\ref{eq:converted optimization problem}) using a dual decomposition method, as the structure of {$y^c_i(t)$} makes it infeasible to decompose the objective function for each RB. In order to overcome this challenge, we convert (\ref{eq:converted optimization problem}) to a decomposable form. Then, we will show that this converted problem is equivalent to (\ref{eq:converted optimization problem}).

For this purpose, the Lagrangian for  problem (\ref{eq:converted optimization problem}) is written as
{
\begin{equation}\label{eq:converted lagrange multipliers}
V\sum_{i,j}p_{ij}(t)+\sum_{i\in \mathcal{M} \cup \mathcal{H}}y^c_i(t)F_i(t)+\sum_{i\in \mathcal{M} \cup \mathcal{H}}\lambda_i \big(a_i(t)-r_i(t)\big),
\end{equation}}
where $\lambda_i$ is the Lagrange multiplier.  As we know, {$y^c_i(t)=e^{-\big(r_i(t)-a_i(t)\big)D_i^{\max}}-\epsilon$}. Therefore, the only decision variables are allocation of resource blocks to the users and allocating power to each RB. Although $F_i(t)$ is a function of {$y^c_i(t)$}, it is not a decision variable and is treated as a constant.  Hence, (\ref{eq:converted lagrange multipliers}) can be rewritten as 
{
\begin{equation}\label{eq:lagrangian function}
V\sum_{i,j}p_{ij}(t)+\sum_{i\in \mathcal{M} \cup \mathcal{H}}e^{-\big(r_i(t)-a_i(t)\big)D_i^{\max}}F_i(t)+\sum_{i\in \mathcal{M} \cup \mathcal{H}}\lambda_i\big(a_i(t)-r_i(t)\big).
\end{equation}}
The main optimization problem consists of two components. First, minimizing the total power of the BS with  weight $V$, and second, minimizing the summation $\sum_{i\in \mathcal{M} \cup \mathcal{H}}e^{-r_i(t)}$ which has a weight $F_i(t)e^{-a_i(t)D_i^{\max}}$ for each user $i$.

As we can see, (\ref{eq:lagrangian function}) is not decomposable for each RB. Here we will have an approximation of (\ref{eq:converted optimization problem}) and then propose an algorithm to solve this approximation efficiently.  In this C-additive approximation, $\sum_{i\in \mathcal{M} \cup \mathcal{H}}e^{-\big(r_i(t)-a_i(t)\big)D_i^{\max}}F_i(t)$ in  (\ref{eq:lagrangian function}) is substituted with its linear approximation of exponential term $e^{-x}$ at $x=0$.
{
\begin{equation}\label{eq:C-additive approximation}
\sum_{i\in \mathcal{M} \cup \mathcal{H}} -\big(r_i(t)-a_i(t)\big)D_i^{\max} F_i(t).
\end{equation}
}

In the original problem, if {$y^c_i(t)$} starts to become greater than zero for user $i$, then $F_i(t)$ will increase and it will give more weight to the term $e^{-\big(r_i(t)-a_i(t)\big)D_i^{\max}}$. As a result, the algorithm allocates more resources to user $i$ such that it minimizes $e^{-\big(r_i(t)-a_i(t)\big)D_i^{\max}}$ for user $i$, and accordingly, {$y^c_i(t)$} decreases. Hence, $F_i(t) e^{-\big(r_i(t)-a_i(t)\big)D_i^{\max}}$ plays the role of feedback in the system. 
As we can see from (\ref{eq:C-additive approximation}), this approximation will not change this feedback mechanism and plays the same role in the system.
Therefore, we can write
{
\begin{align}
&\underset{\boldsymbol{P}, \mathbf{\rho}}{\min}\big\{V\sum_{i,j}p_{ij}(t)+\sum_{i\in \mathcal{M} \cup \mathcal{H}} -\big(r_i(t)-a_i(t)\big)D_i^{\max}F_i(t) \big\}\nonumber\\
&<C+\underset{\boldsymbol{P}, \mathbf{\rho}}{\min}\big\{V\sum_{i,j}p_{ij}(t)+\sum_{i\in \mathcal{M} \cup \mathcal{H}}e^{-\big(r_i(t)-a_i(t)\big)D_i^{\max}}F_i(t)\big\}.
\end{align}}
Using this C-additive approximation, it can be easily proved that all terms are mean-rate stable. Hence, (\ref{eq:constrain1}) in the original problem is satisfied \cite{neely2010stochastic}. 
Finally, problem (\ref{eq:OrigOptim}) can be presented as:
{
\begin{subequations}\label{eq:final optimization problem}
	\begin{align}
	\min_{\rhob(t),\Pb(t)} \qquad &V\sum_{i,j}p_{ij}(t)-\sum_{i\in \mathcal{M} \cup \mathcal{H}} \big(r_i(t)-a_i(t)\big)D_i^{\max}F_i(t), \nonumber\\
	\text{s.t.} \qquad &r_i(t)>a_i(t), \label{eq:minimum rate lambda}\\
	&p_{ij}(t) \geq 0, \quad \forall i \in \mathcal{H} \cup \mathcal{M}, j\in \mathcal{K},\\
	&\rho_{ij}(t) \in \{0,1\}, \quad  \forall i \in \mathcal{H} \cup \mathcal{M}, j\in \mathcal{K}, \\
    &\sum_{i \in \mathcal{H} \cup \mathcal{M}}\rho_{ij}(t)=1, \quad\forall j\in \mathcal{K}.\label{eq:coupling constraint}
	\end{align}
\end{subequations}
}
In order to solve this problem, we can decompose it into $K$ subproblems. Since these subproblems are coupled through constraint (\ref{eq:coupling constraint}),  we use the dual decomposition method for solving (\ref{eq:final optimization problem}) \cite{seong2006optimal}.  First, the Lagrangian is written for problem (\ref{eq:final optimization problem}), and in the second step, it is decomposed for each RB. Then, the resource block allocation and the power of each RB are found in terms of the Lagrange multiplier {vector} $\lambdab$. Finally, $\lambdab$ is calculated using an ellipsoid method.

The Lagrangian for problem (\ref{eq:final optimization problem}) is
\begin{align}
\mathcal{L}_a(\boldsymbol{P}, \mathbf{\rho},\lambdab)&=V\sum_{i,j}p_{i,j}(t)+\sum_{i\in \mathcal{M} \cup \mathcal{H}} -\big(r_i(t)-a_i(t)\big)D_i^{\max}F_i(t)-\lambda_i\big(r_i(t)-a_i(t)\big)\nonumber\\
&=V\sum_{i,j}p_{i,j}(t)-\sum_{i\in \mathcal{M} \cup \mathcal{H}} \big(\lambda_i+D_i^{\max}F_i(t)\big) \big(r_i(t)-a_i(t)\big).\label{eq:lagrangian}
\end{align}
One major difference between our problem and conventional power minimization problems is that there is an additional term  $D_i^{\max} F_i(t)$ added to the Lagrange multiplier (the shadow price).
 
In this problem, $D_i^{\max} F_i(t)$ plays the role of a bias term. Therefore, a new hypothetical Lagrange multiplier $\lambda_i'$ is assumed and defined as
$
\lambda_i'=\lambda_i+D_i^{\max}F_i(t).
$
This means that adding  constraint (\ref{eq:constrain1}) to the problem instead of constraint (\ref{eq:minimum rate lambda}) increases the shadow price by a factor of $D_i^{\max}F_i(t)$. Increasing the shadow price for a constraint makes it looser. As a result, in many time slots,  constraint (\ref{eq:minimum rate lambda}) will not be a tight constraint and  the Lagrange multiplier will be set to  $\lambda_i=\big[\lambda_i'-D_i^{\max}F_i(t)\big]^+$. 
the Lagrange dual function is
\begin{equation}\label{eq:lagrange dual problem}
g(\lambdab)=\min_{\rhob(t),\Pb(t)} \quad \mathcal{L}_a(\boldsymbol{P}, \boldsymbol{\rho},\lambdab).
\end{equation}
The minimization problem (\ref{eq:lagrange dual problem}) can be decomposed to $K$ subproblems. $g_j'(\lambdab)$ can be written as
\begin{equation}
g_j'(\lambdab)=\min_{\Pb(t)} \quad V\sum_i p_{i,j}-\sum_{i\in \mathcal{M} \cup \mathcal{H}} \big(\lambda_i+D_i^{\max}F_i(t)\big) \big(W \log_2(1+K \frac{h_{i,j}p_{i,j}}{\sigma^2})\big),
\end{equation}
where $\mathcal{D}$ is a set of feasible $p_{ij}$s in which for RB $j$, there is only one $i$ that $p_{ij}\neq0$. Hence, $g(\lambdab)$ is
\begin{equation}\label{eq:dual_objective}
g(\lambdab)=\sum_j g_j'(\lambdab)+\sum_{i\in \mathcal{M} \cup \mathcal{H}} \big(\lambda_i+D_i^{\max}F_i(t)\big) \big(a_i(t)\big).
\end{equation}
If $\lambdab$ is fixed, $g_j'(\lambdab)$ is a convex function of $\boldsymbol{P}$. Therefore,  $\boldsymbol{P}$ is found by taking a derivate with respect to $p_{ij}$ and setting it to zero. This results in 
\begin{equation}\label{eq:pijfind}
p_{ij}=\Big[\frac{\big(\lambda_i+D_i^{\max}F_i(t)\big) W}{V \log_2 }-\frac{\sigma^2}{K h_{ij}}\Big]^+.
\end{equation}

The optimal RB allocation for RB $j$ is $k(j)$, and can be written as
\begin{align}
k(j)=\underset{i}{\text{argmin}} V\sum_i p_{i,j}-\sum_{i\in \mathcal{M} \cup \mathcal{H}} \big(\lambda_i+D_i^{\max}F_i(t)\big) \big(W \log_2(1+K \frac{h_{i,j}p_{i,j}}{\sigma^2})\big), \label{eq:find the subchannel assignment}\\
g_j'(\lambdab)=\underset{i}{\min} V\sum_i p_{i,j}-\sum_{i\in \mathcal{M} \cup \mathcal{H}} \big(\lambda_i+D_i^{\max}F_i(t)\big) \big(W \log_2(1+K \frac{h_{i,j}p_{i,j}}{\sigma^2})\big).
\end{align}
Thus, $\rho^*_{ij}$ and $p^*_{ij}$ will be given by: 
\begin{align}
&\rho^*_{ij}=\begin{cases}
1, &i=k(j),\\
0, &\text{otherwise}.
\end{cases}
& p^*_{ij}=\begin{cases}
p_{ij}, &i=k(j),\\
0, &\text{otherwise}.
\end{cases}
\end{align}

\begin{algorithm}[!t]
 \caption{{Resource allocation algorithm}}
 {\begin{algorithmic}[1]\label{alg:RA}
  \STATE Obtain $D_i^{\max(t)},\, \epsilon_i(t)\, \forall i \in \mathcal{H} \cup \mathcal{M}$ using PDI algorithm (Algorithm 2).
  \STATE Find $F_i(t),\, \forall i \in \mathcal{H} \cup \mathcal{M},$ using (\ref{eq:F_iEq})
  \STATE Initialize $\lambdab$
  \WHILE{convergence condition is not satisfied} 
    \STATE Find $p_{ij},\, \forall i \in \mathcal{H} \cup \mathcal{M}, j\in \mathcal{K},$ using the updated $\lambdab$ (\ref{eq:pijfind})
  \STATE for each RB $j\in \mathcal{K}$, find $k(j)$ by searching over all users $i \in \mathcal{H} \cup \mathcal{M}$ using (\ref{eq:find the subchannel assignment}) and then assign $\rho^*_{ij}$ and $p^*_{ij}$ for all $ i \in \mathcal{H} \cup \mathcal{M}, j\in \mathcal{K},$
  \STATE Use the ellipsoid method to find $\lambdab$
  \ENDWHILE
 \end{algorithmic} }
 \end{algorithm}
 
Hence, the optimal rate becomes $r^*_i= \sum_j W \log_2(1+K \frac{h_{i,j}p^*_{i,j}}{\sigma^2})$. The only parameter that affects this joint RB and power allocation is $\lambdab$. As the number of RBs increases, the duality gap in this problem approaches zero \cite{seong2006optimal}. We know that the optimal value is found by maximization of $g(\lambdab)$ with respect to $\lambdab$. In order to find $\lambdab$,  we use the ellipsoid method \cite{yu2006dual}, and to do so, we have to find the sub-gradient for the dual objective $g(\lambdab)$.  The following theorem will show that the subgradient for (\ref{eq:lagrangian}) is a vector with elements $\zeta_i=a_i-r_i$. 
\begin{theorem}
The subgradient of the dual optimization problem with dual objective defined in (\ref{eq:dual_objective}), is the vector $\boldsymbol d$ whose elements $\zeta_i, \, \forall i \in \mathcal{H} \cup \mathcal{M}$ are given by: 
\begin{equation}
\zeta_i=\begin{cases}
a_i-r_i^*, &a_i \geq r_i^*,\\
0, &a_i<r_i^*.
\end{cases}
\end{equation}
\end{theorem}
\begin{IEEEproof}
Since 
\begin{equation}
g(\lambdab)=\min_{\Pb,\rho}\mathcal{L}_a(\boldsymbol{P}, \boldsymbol{\rho},\lambdab)=\mathcal{L}_a(\boldsymbol{P}^*, \mathbf{\rho}^*,\lambdab),
\end{equation}
 we have:
\begin{align}
g(\boldsymbol{\deltab} )\leq& \mathcal{L}_a(\boldsymbol{P}^*, \mathbf{\rho}^*,\deltab)\nonumber\\
=&V\sum_{i,j}p^*_{i,j}(t)-\sum_{i\in \mathcal{M} \cup \mathcal{H}} \big(\delta_i+D_i^{\max}F_i(t)\big) \big(r^*_i(t)-a_i(t)\big)\nonumber\\
=&V\sum_{i,j}p^*_{i,j}(t)-\sum_{i\in \mathcal{M} \cup \mathcal{H}} \big(\lambda_i+D_i^{\max}F_i(t)\big) \big(r^*_i(t)-a_i(t)\big)\nonumber\\
&+(\lambda_i-\delta_i)\big(r^*_i(t)-a_i(t)\big)=g(\lambdab)+(\lambdab-\deltab)^T\boldsymbol \zeta',
\end{align}
where 
$
\boldsymbol \zeta'=\begin{bmatrix}
r_1^*-a_1 &\cdots & r_{N+M}^*-a_{N+M}
\end{bmatrix}^T.
$

{However, because of the term $D_i^{\max}F_i(t)$,  when $\lambda_i=0$ and $a_i<r_i^*$, the direction of $\boldsymbol \zeta'$ will be infeasible}. Using the projected subgradient method \cite{boyd2006subgradient}, we can transform this infeasible direction to a feasible one. The update rule for projected subgradient is:
$
\lambdab^{(k+1)}=\Pi (\lambdab^{(k)}- \alpha_k \zeta_k')
$
where $\alpha_k$ is the step size and $\Pi$ is the Euclidan projection on the feasible set. Since the feasible set is $\lambda_i>0$, we can see that 
\begin{equation}
\Pi (\lambdab^{(k)}- \alpha_k \zeta_k')=\lambdab^{(k)}- \alpha_k \zeta_k,
\end{equation}
where: 
\begin{equation}
\zeta_i=\begin{cases}
\zeta_i', &\zeta_i'\geq0\\
0, &\zeta_i'<0
\end{cases}=\begin{cases}
a_i-r_i^*, &a_i \geq r_i^*,\\
0, &a_i<r_i^*.
\end{cases}
\end{equation}
\end{IEEEproof}
{Algorithm \ref{alg:RA} summarizes our proposed resource allocation algorithm.} 

\subsection{Complexity Analysis}
Next, we find the complexity of our algorithm which needs to be run in each iteration. There are $K$ RBs in our problem, for each of which (\ref{eq:find the subchannel assignment}) needs to be evaluated for $M+N$ users. It takes $\mathcal{O}\big((M+N)K\big)$ times to solve a primal problem.  Subsequently, the dual problem will be solved,  which gives us the optimal value of $\lambdab$ in an $M+N$ dimensional space and has a complexity of $\mathcal{O}\big((M+N)^2\big)$. Therefore, the overall complexity should be $\mathcal{O}\big((M+N)^3K\big)$. However, as mentioned before, adding $D_i^{\max}F_i(t)$ to the Lagrange multiplier sets a major part of it to zero, and as a result, the order of complexity will decrease to $\mathcal{O}\big((M+N)K\big)$. 
Given the low-complexity of the proposed algorithm, in practice, it can be easily run periodically by the network at each time slot $t$, so as to effectively adapt to dynamic, time-varying changes in both the human delay perceptions of the users and the wireless channel.

	\section{Simulation Results and Analysis}\label{sec:Simul}
\begin{figure}[!t]
    \begin{minipage}{0.5\textwidth}
		\centering
		\includegraphics[scale=.6,trim={0.6cm 0 0 0}]{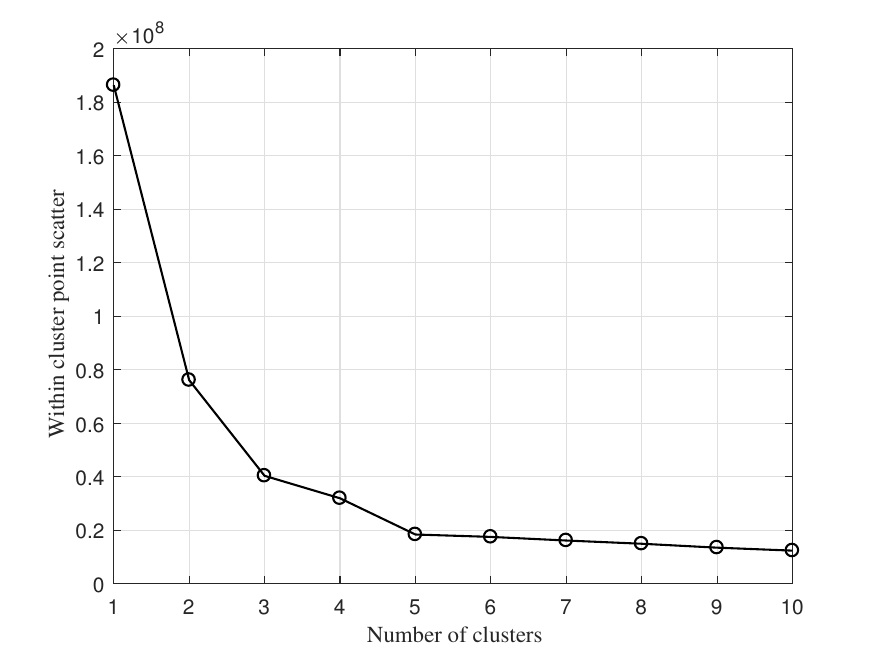}
		\caption{ٍWithin point scatter for the EM clustering method on the datasest.}
		\label{fig:WPS}
	\end{minipage}
    	\begin{minipage}{0.5\textwidth}
		\centering
		\includegraphics[scale=.6,trim={0.6cm 0 0 0}]{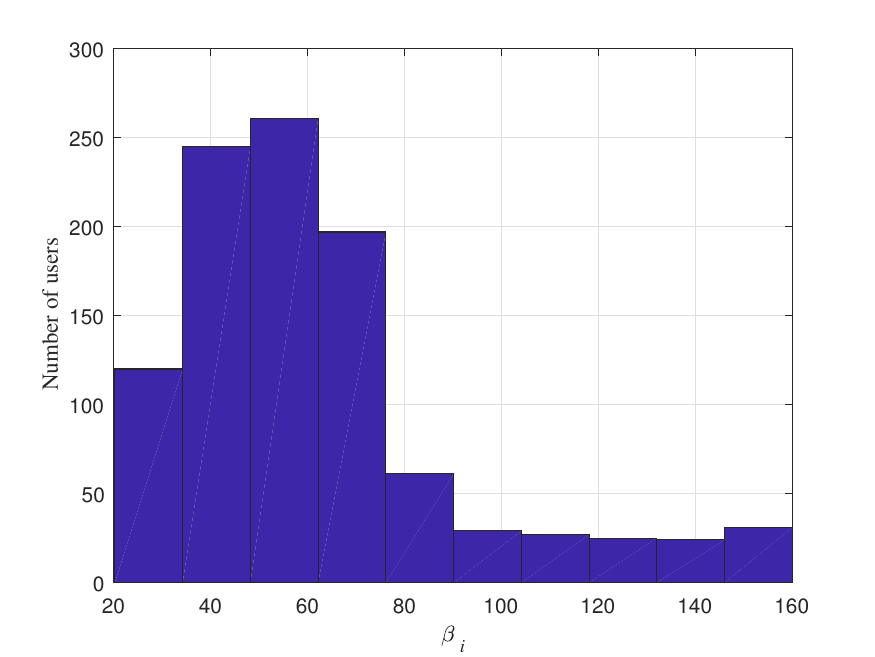}
		\caption{ٍDistribution of $\beta_i(t)$ for the 1000 users in dataset.}
		\label{fig:Hist}
        \end{minipage}
	\end{figure}

	For our simulations, we consider the dataset in \cite{yang2015prospect} to model the delay perception of a human user. In \cite{yang2015prospect}, the authors conducted human subject studies using 30 human users, where each subject is asked to rate the quality of 5 movies while the delay and packet loss in the system is being increased. We used the average score of each human user to estimate their delay perception. {In \cite{yang2015prospect}, The highest delay that the test subject is not able to sense is considered as the delay perception of this subject, and, this, matches our definition of delay perception.} We also used a variation of the bootstrap method \cite{friedman2001elements} to increase the number of data points to 1000. We can see the histogram of the delay perception for these 1000 data points in Fig. \ref{fig:Hist}.

{To the best of our knowledge, no dataset which includes features for each human user as well as human delay perception currently exists. Hence, we attribute three continuous features to each user. The process of adding features starts by clustering the delay perceptions $\beta_i(t)$ for 1000 users. Then, we choose a random mean vector and a random positive semidefinite covariance matrix for each cluster and use them to create multivariate random Gaussian features for each data in the cluster. In consequence the random features have: 1) a GMM structure and 2) a predictive ability for $\beta_i(t)$.} Hence, each user is associated with a vector $\wb \in \mathds{R}^4$.

We consider a network with a bandwidth of $10$~MHz, $a_i(t)=1$~Mbps, $\sigma^2=-173.9$~dBm, and $\epsilon=0.05$.  We use a circular cell with the cell radius of $1.5$~km. We set the path loss exponent to $3$ (urban area) and the carrier frequency to $900$~MHz. The packet length is an exponential random variable with an average size of $10$~kbits. We use 5 MTD and 5 UE in the system  and we set $D_i^{\max}$ to $20$~ms for them, unless otherwise mentioned. For the brain aware users, we arbitrarily select 5 UE in the system out of all data points. {The brain-unaware case is QoS-aware and power-aware. In this case, $D_i^{\max}$ and $\epsilon$ are not functions of $\beta_i(t)$.}

Fig. \ref{fig:WPS} shows the within cluster point scatter for the EM algorithm in our dataset. This \emph{within cluster point scatter} for a clustering $C$ is defined as \cite{friedman2001elements}: 
$
W(C)=\frac{1}{2}\sum_{k=1}^n\sum_{c(i)=k}\sum_{c(i')=k}d(x_i,x_{i'}),
$
where $d$ is an arbitrary distance metric. In essence, the within cluster point scatter is a loss function that allows the determination of hyper-parameters in the clustering algorithm. The hyper-parameter that we seek to find here is the number of clusters in the dataset. As we can see from Fig. \ref{fig:WPS}, after the number of clusters reaches $5$, increasing the number of clusters does not decrease the within cluster point scatter substantially. Hence, the optimal number of clusters is $5$. {This method of model selection known as elbow method allows the algorithm to avoid overfitting.}

\begin{figure}[!t]
    \begin{minipage}{0.5\textwidth}
		\centering
		\includegraphics[scale=.6,trim={0.6cm 0 0 0}]{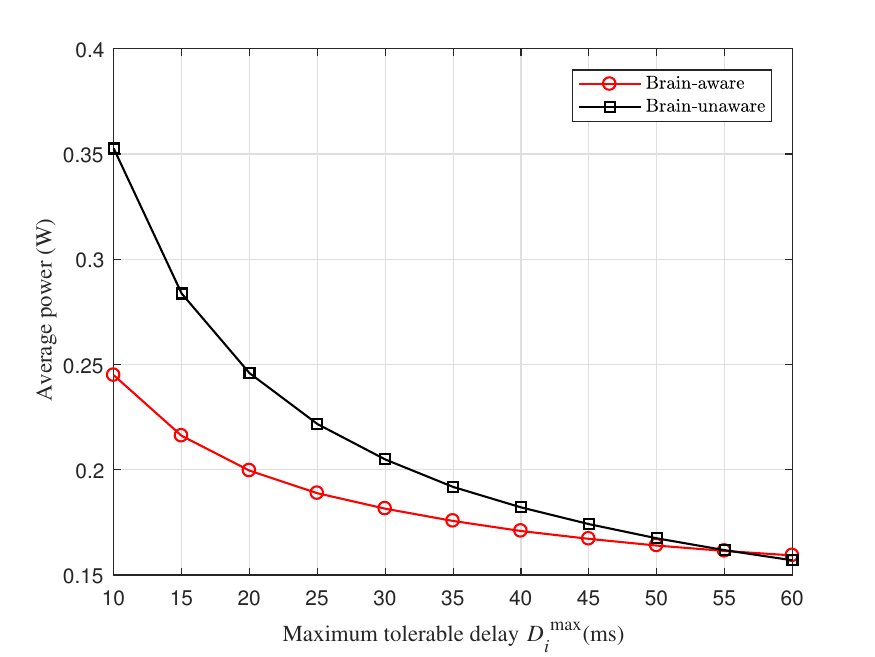}
		\caption{ٍAverage power usage of the system as function of different latency requirements for the users. }
		\label{fig:power versus delay}
	\end{minipage}
    	\begin{minipage}{0.5\textwidth}
		\centering
		\includegraphics[scale=.6,trim={0.6cm 0 0 0}]{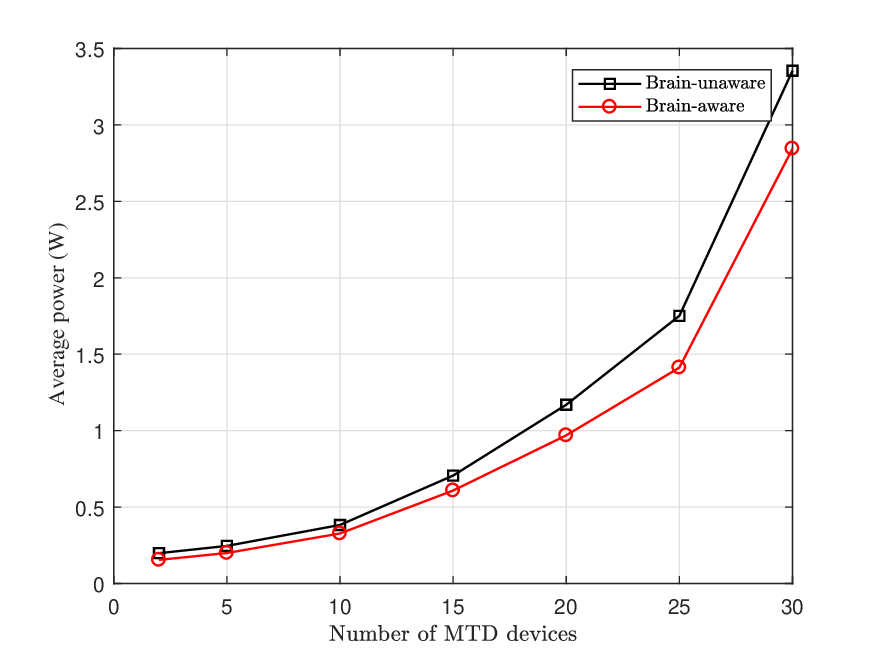}
		\caption{ٍ{Average power usage of the system for different number of MTDs and 5 UEs with  $D_i^{\max}=20$~ms.}}
		\label{fig:sweepMTD}
        \end{minipage}
	\end{figure}

Fig. \ref{fig:power versus delay} shows the total BS power resulting from the proposed brain-aware case and from a brain-unaware case in which  UEs have a fixed constraint (\ref{eq:constrain1}) with $D_i^{\max}$ between $10$~ms to $60$~ms. Here, the total power is the objective of main optimization problem (\ref{eq:OrigOptim}). 
	Fig. \ref{fig:power versus delay} shows that, as the latency increases, the total power decreases, because it is easier to satisfy  constraint (\ref{eq:constrain1}) at higher latencies. 
	Also, at higher delays, being brain-aware will no longer yield substantial gains, since $\beta_i(t)$ and $D_i^{\max}$ become close to each other and learning $\beta_i(t)$ cannot save resources for the system. In contrast, in Fig. \ref{fig:power versus delay}, we can see that for stringent low-latency requirements, the proposed brain-aware approach yields significant gains in terms of saving power. In particular, for $10$~ms delay in (\ref{eq:constrain1}), Fig. \ref{fig:power versus delay} shows that the BS in brain-unaware approach uses $44$~\% more power compared to the brain-aware case. These results stem from the fact that a brain-aware approach can minimize waste of resources and provide service to the users more precisely based on their real brain processing power. 
	Fig. \ref{fig:sweepMTD} shows average BS power for different number of MTDs. As we can see from Fig. \ref{fig:sweepMTD}, the brain-aware approach will always outperform the brain-unaware approach as the number of MTD increases. For the case of 30 MTD user, the BS in brain-unaware approach uses $16$\% more power compared to the brain-aware case. This is due to fact that brain-aware approach can allocate resources more efficiently in case of a shortage in resources.

		\begin{figure}[!t]
    \begin{minipage}{0.5\textwidth}
    	\centering
		\includegraphics[scale=.6,trim={0.6cm 0 0 0}]{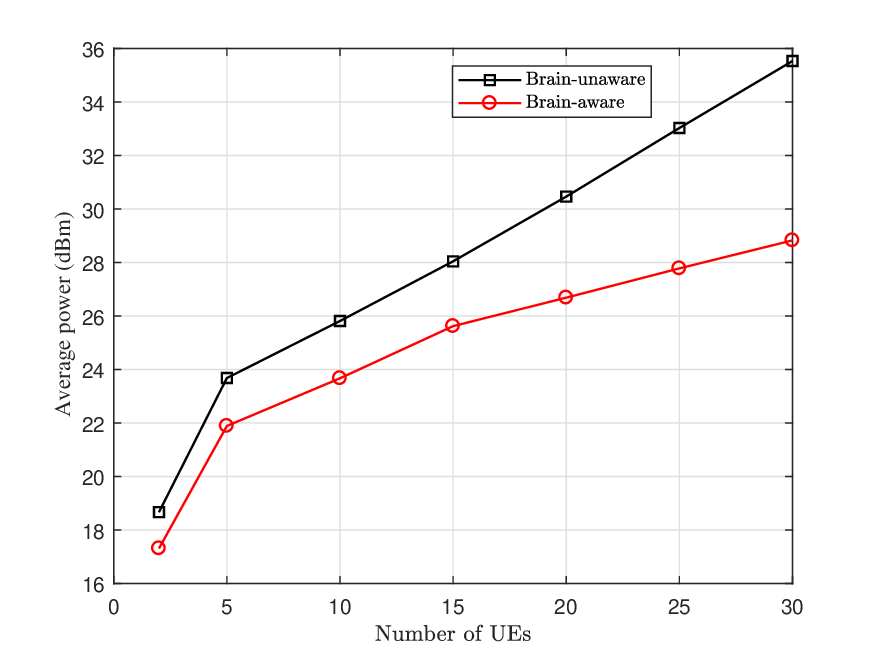}
		\caption{ٍ{Average power usage of the system for different number of UEs with  $D_i^{\max}=20$~ms.}}
		\label{fig:sweepuser}
   \end{minipage}
    	\begin{minipage}{0.5\textwidth}
	\centering
	\includegraphics[scale=.6]{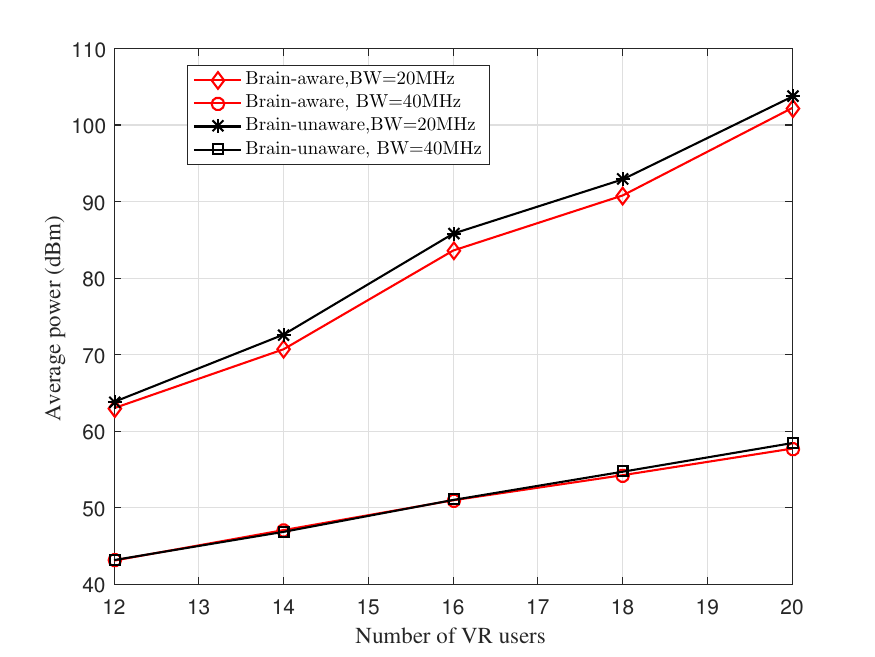}
	\caption{{The effect of number of VR users with the rate of 25.31~Mbps and $D_i^{\max}=20$~ms on the power usage of the system.}}
	\label{fig:VR}
        \end{minipage}

	\end{figure}

In Fig. \ref{fig:sweepuser}, we show the average power usage of the system when the number of UEs increases from $2$ to $30$ with $D_i^{\max}$ set to $20$~ms. As the number of users increases, the average power consumption of the system will also increase. This is due to the fact that increasing the number of users will decrease the bandwidth per user. Since the delay and rate requirements of each user are still unchanged, the system needs to use more power to compensate for the bandwidth deficiency. From Fig. \ref{fig:sweepuser}, we can see that, in the case of $30$ users, the brain-aware system is able to save  $6.7$~dB ($78$\%) on average in the BS power. The brain-aware system can allocate resources based on each user's actual requirement  instead of the predefined metrics and this leads to this significant saving in the power consumption of the BS.

	\begin{figure}[!t]
    \begin{minipage}{0.5\textwidth}
		\centering
		\includegraphics[scale=.6,trim={0.6cm 0 0 0}]{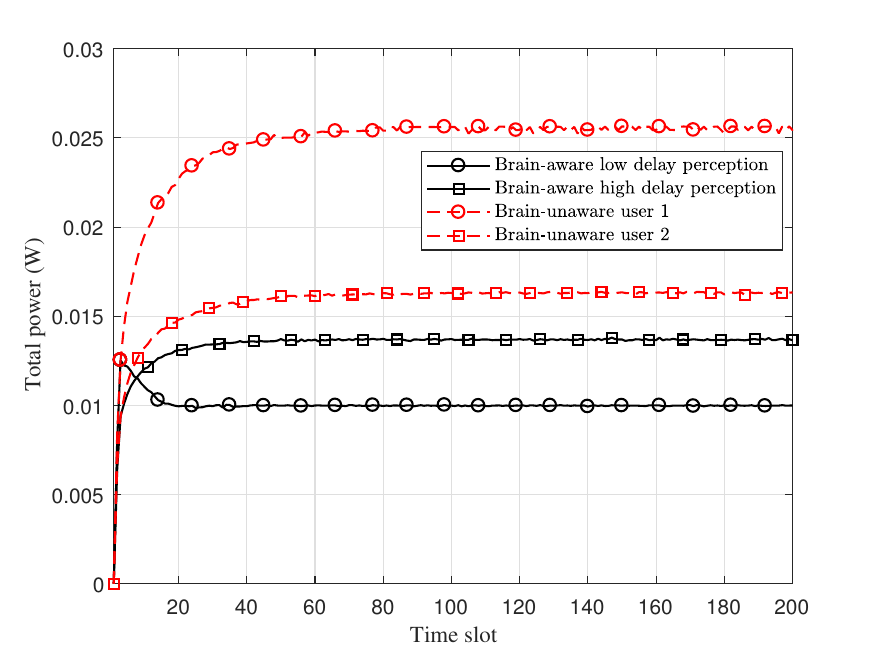}
		\caption{{Transmit power for $4$ different users. The delay perception of two of the users is learned. Low and high delay perception users have delay perception of $26.8$~ms and $133.73$~ms, respectively.}}
		\label{fig:youngOldpower}
	\end{minipage}
    \begin{minipage}{0.5\textwidth}
		\centering
		\includegraphics[scale=.6,trim={0.6cm 0 0 0}]{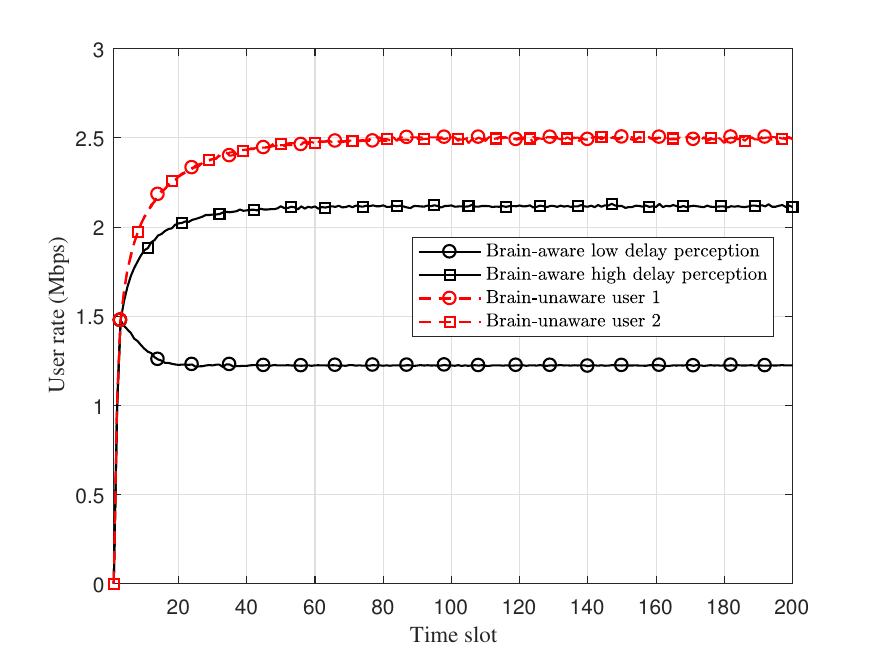}
		\caption{{Transmission rate for four different users. The delay perception of two of the users is learned. Low and high delay perception users have delay perceptions of $26.8$~ms and $133.73$~ms, respectively.}}
		\label{fig:YoungOldRate}
        \end{minipage}
	\end{figure}

{In Fig. \ref{fig:VR}, we show the average power consumed in the system for different number of virtual reality (VR) users. For the VR simulations, we assumed an arrival rate of 25.31~Mbps for each user \cite{Mingzhe2017Globecom} and have used bandwidths of 20~MHz and 40~MHz. We can see that, the system is able to save power up to 40\%  and 15\% compared to the brain-unaware scenario in the case of 20 MHz, and 40 MHz bandwidth, respectively. Fig. \ref{fig:VR} also shows that the proposed approach is able to allocate resources more efficiently when resources are scarce, i.e. in the 20 MHz case. Also, we can see that increasing the bandwidth will decrease the total power usage in the system which is an inherent feature of communication systems.}

\begin{figure}[!t]
    \begin{minipage}{0.5\textwidth}
		\centering
		\includegraphics[scale=.6,trim={0.6cm 0 0 0}]{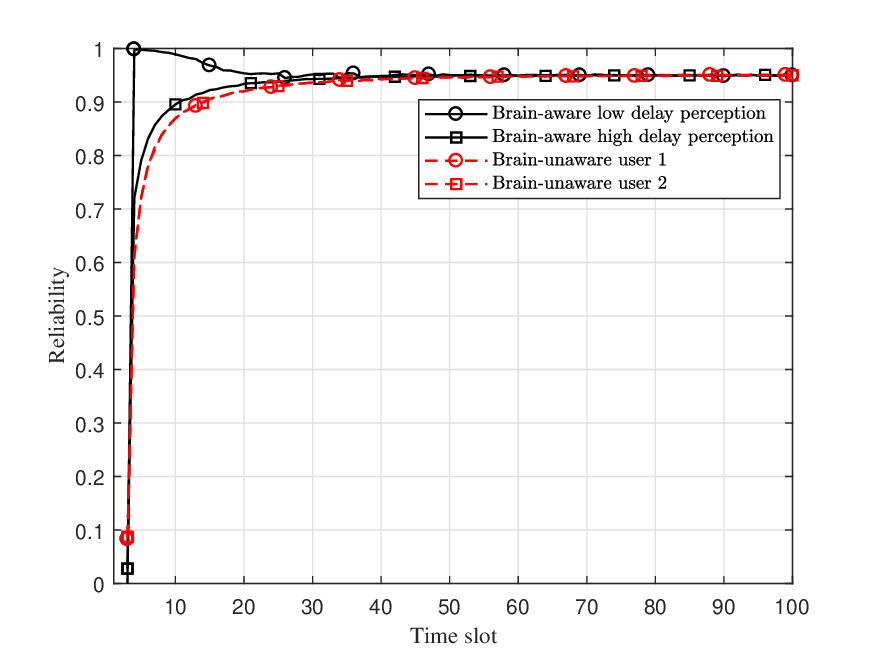}
		\caption{Reliability for {4} different users. The delay perception of two of the users is learned.}
		\label{fig:YoungoldRely}
	\end{minipage}
\begin{minipage}{0.5\textwidth}
		\centering
		\includegraphics[scale=.6,trim={0.6cm 0 0 0}]{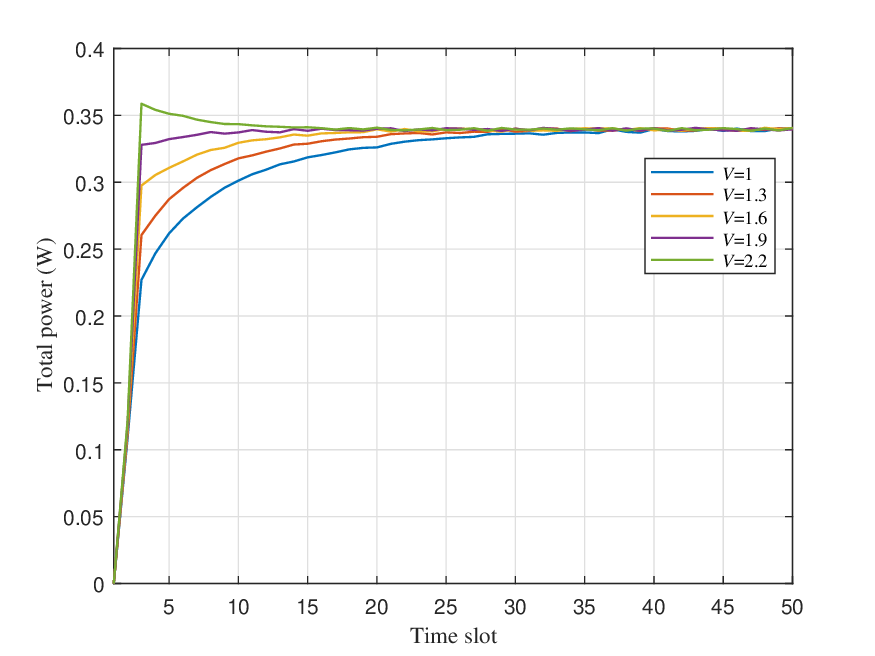}
		\caption{{Effect of balancing parameter $V$ in (\ref{eq:converted cost func}) on the convergence of the resource allocation algorithm.}}
		\label{fig:V}
        \end{minipage}
	\end{figure}

In Fig. \ref{fig:youngOldpower}, Fig. \ref{fig:YoungOldRate}, and Fig. \ref{fig:YoungoldRely}, we consider the case of 7 UEs and 5 MTDs. Two UEs are chosen as brain-aware users and their delay perception is learned by the PDI method. One of the brain-aware UEs has a delay perception of $\beta_i(t)=133.73$~ms, and the other one has $\beta_i(t)$ equal to $26.8$~ms. The system does not learn the delay perception  of the 5 remaining UEs and, hence, it allocates resources to them by using a predefined delay requirement (brain-unaware users).

As we can see in Fig. \ref{fig:youngOldpower}, the power consumption of the first two brain-aware users will be less than that of the brain-unaware users. Moreover, the power consumption for a user with higher delay perception will be less than that of a user with lower delay perception. This shows that the system can successfully allocate resources according to the delay perception of the users. Furthermore, the power consumption related to each user with predetermined delay requirements is different, due to their different channel gains. However, as we will see later, the system is robust to such differences and can guarantee the reliability and rate requirements for users having different channel gains.

In Fig. \ref{fig:YoungOldRate}, we show the transmission rate for four different users. We can see that the rate for brain-unaware users with predetermined delay will converge to  $2.5$~Mbps. This rate will ensure the reliability for these users. However, the rate of the users with learned delay perception will converge to a smaller rate. This is due to the fact that these users' actual requirements are known to the system, and the system uses this knowledge to avoid unnecessarily wasting resources. However, as we will see next, this rate reduction does not change the reliability for these users.

Fig. \ref{fig:YoungoldRely} shows the reliability for the four aforementioned users. As we can see, the reliability of all the users will converge to $95$~\%, which is the target reliability value for the users. We can see that the system is able to ensure reliability for the users with identified delay perceptions as well as the users with predefined delay requirements. {However, as observed from Fig. \ref{fig:youngOldpower}, the system uses $45$\% less power  for those users for which the delay perception is learned. } 

Finally, Fig. \ref{fig:V} investigates the effect of parameter $V$ for the system with 5 MTDs and 5 UEs. We can see that, as $V$ increases from $1$ to $1.9$, the convergence time decreases from $40$ iterations to $15$ iterations. Nevertheless, increasing $V$ will make the algorithm unstable, and as we can see, increasing it to $2.2$ will create an overshoot which is $11$\% higher than the final value. Hence, parameter $V$, if adjusted correctly, can create a balance between stability and  convergence rate  of our algorithm.      
	\section{Conclusion}
	
	In this paper, we have introduced and formulated the notion of delay perception of a human brain, in wireless networks with humans-in-the-loop. Using this notion, we have defined the concept of effective delay of human brain. { To quantify this effective delay, we have developed a  learning method, named PDI, which consists of an unsupervised and supervised learning part.} We have then shown that PDI can predict the effective delay for the human users and find the  reliability of this prediction. Then, we have derived a closed-form relationship between the reliability measure and wireless physical layer metrics. Next, using this relationship and the PDI method, we have proposed a novel approach based on Lyapunov optimization for allocating radio resources to human users while considering the reliability of both machine type devices and human users. 
	Our results have shown that the proposed brain-aware approach can save a significant amount of power in the system, particularly for low-latency applications and congested networks. To our best knowledge, this is the first study on the effect of human brain limitations in wireless network design.
This paper only scratched the surface of an emerging research area that admits several future extensions. On the one hand, we can extend the studied framework to accommodate other brain-related features beyond the mode of the brain. Examples of such features include perceptual memory and consistency constraints. On the other hand, we can develop recurrent neural network models to capture how the sequence in the brain mode can dynamically change. Finally, another important future work is to conduct real-world experiments with actual users to gather empirical date on brain behavior so as to refine the developed solution.

	\appendices
	
	\section{Proof of Theorem \ref{lemma:confidence region single mode}}
	We assume that a single brain mode is dominant for each user at each time. We index this single mode as $k$. For each user $i$ with this dominant mode, {$\wb_i=[w_1,\cdots,w_{d+1}]$} has the following probability density function:
	{\begin{equation}
	p(\wb_i)=	|2\pi \Sigmab_k|^{-\frac{1}{2}} \exp\Big[-\frac{1}{2}(\wb_i-\mub_k)^T\Sigmab_k^{-1}(\wb_i-\mub_k)\Big].
	\end{equation}  }
	We want to find the smallest region $\mathcal{D}$ in $\mathds{R}^{d+1}$, in which the delay perception lies  with probability $\gamma$, i.e., 
	{\begin{equation}
	\label{eq:Region D definition}
	\int \cdots \int_\mathcal{D}\, p(w_1,w_2,\ldots,w_{d+1}) \,dw_1 \!\cdots dw_{d+1}=\gamma.
	\end{equation} }
	$\Db$ is not a unique region. However, the objective is to find the smallest region.
	To this end, we need to find the region where {$f(w_1,w_2,\ldots,w_{d+1})$} has the greatest value, i.e., if
	{\begin{equation}
	\label{eq:appendix:equal to}
	\int \cdots \int_{\Db_1}\, p(w_1,w_2,\ldots,w_{d+1}) \,dw_1 \!\cdots dw_n=\int \cdots \int_{\Db_2}\, p(y_1,y_2,\ldots,y_n) \,dy_1 \!\cdots dy_n,
	\end{equation} }
	and also{
	\begin{equation}
	\label{eq:appendix:less than}
	p(y_1,y_2,\ldots,y_{d+1}) \leq p(w_1,w_2,\ldots,w_{d+1}) \qquad\forall \boldsymbol{y} \in \Db_2,\,\, \forall \wb_i \in \Db_1,
	\end{equation}}
	then 
	{\begin{equation}
	\int \cdots \int_{\Db_1} \,dw_1 \!\cdots dw_{d+1} \leq \int \cdots \int_{\Db_2} \,dy_1 \!\cdots dy_{d+1},
	\end{equation}}
	which implies that the volume of the region $\Db_1$ is smaller than the volume of $\Db_2$. Hence, if we find the region $\Db$ for which (\ref{eq:Region D definition}) holds, and, using (\ref{eq:appendix:less than}), show that all other regions for which (\ref{eq:Region D definition}) holds have greater volumes, then, we would have found the smallest region $\Db$, in which the human behavior will stay with the probability $\gamma$.

	Since $\wb_i$ is distributed according to a multivariate Gaussian, we can find the region where it has the highest probability density, i.e.,  {$\big\{\wb_i|p(\wb_i)>C_1\big\}$}. This region can be written as:
	\begin{equation}
	\bigg\{\wb_i \bigg|	\,|2\pi \Sigmab_k|^{-\frac{1}{2}} \exp\Big[-\frac{1}{2}(\wb_i-\mub_k)^T\Sigmab_k^{-1}(\wb_i-\mub_k)\Big]>C_1 \bigg\},
	\end{equation}
	which is equivalent to
	\begin{equation}
	\label{eq:appendix:ellipsoid}
	\Db=\Big\{\wb_i\Big|(\wb_i-\mub_k)^T\Sigmab_k^{-1}(\wb_i-\mub_k)<C_2\Big\},
	\end{equation}
	where $C_2$ is a positive constant and equals  $-\ln |2 \pi\Sigmab_k|^{\frac{1}{2}} C_1$. Since $\Sigmab_k$ is a positive definite matrix, (\ref{eq:appendix:ellipsoid}) is the inner volume of an ellipsoid in a $d$ dimensional space. 
	
	We now conjecture that this ellipsoid $\Db$ is the smallest region, in which the delay perception lies with probability $\gamma$, i.e., the probability of $\wb_i$ being in this region is $\gamma$. We use a proof by contradiction to show this. Consider that there exists any other space $\Eb$ which is smaller than $\Db$, and the probability of $\wb_i$ being in this region is $\gamma$. We can partition $\Eb$ into two parts $\Ab=\Eb \cap \Db$ and $\Eb_2=\Eb \cap \Db'$, where $\Db'$ is the complement of the set $\Db$. We also define $\Db_2=\Db \cap \Eb'$. We know that
	{\begin{align}
	&\int_\Db p(\wb_i) d\wb_i=\int_{\Ab} p(\wb_i) d\wb_i+\int_{\Db_2} p(\wb_i) d\wb_i \\
	=&\int_\Eb p(\wb_i) d\wb_i=\int_{\Ab} p(\wb_i) d\wb_i+\int_{\Eb_2} p(\wb_i) d\wb_i	=\gamma.
	\end{align}}
	Hence, $\int_{\Db_2} f(\wb_i) d\wb_i=\int_{\Eb_2} f(\wb_i) d\wb_i$.  Since
	{\begin{equation}
	p(\wb_i)<C_1\leq p(\boldsymbol{y}) \qquad \forall \wb_i \in \Eb_2, \boldsymbol{y} \in \Db_2,
	\end{equation} }
	using (\ref{eq:appendix:equal to}) and (\ref{eq:appendix:less than}) we have 
	$
	\int_{\Eb_2}  d\wb_i<\int_{\Db_2}  d\wb_i.
	$
	This means that the set $\Eb$ has a bigger volume than $\Db$, which is a contradiction to our first assumption.  This proves that region $\Db$ is the smallest region in $\mathds{R}^{d+1}$ that has the probability $\gamma$.
	
	Next, we find the relation between $C_2$ and $\gamma$.  
	$\gamma$ can be defined as {$\int_\Db p(\wb_i) d\wb_i$} and can be calculated using chi-square distribution \cite{berger1980robust}. The region $\Db$ can be written as
	
	\begin{equation}
	D=\big\{\wb_i|(\wb_i-\mub_k)^T\Sigmab_k^{-1}(\wb_i-\mub_k)\leq Q_{d+1}(\gamma)\big\},
	\end{equation}
	where $Q_{d+1}(\gamma)$ is the quantile function of the chi-square distribution with $d+1$ degrees of freedom. It is defined as
	{$
	Q_{d+1}(\gamma)=\inf\Big\{x\in \mathds{R}| \gamma\leq \int_{0}^{x} \chi_d^2(u) du\Big\}.
	$}
	
	Having defined the confidence region based on $\gamma$, we now must find the edges of this ellipsoid. We know that the center of this ellipsoid is $\mub_k$. We need to solve the  following optimization problem:
\begin{equation}
	\underset{\wb_i}{\min} \,\,\text{or}\,\, \underset{\wb_i}{\max} \,\,\eb_j^T\wb_i,\quad \text{subject to} \,\,  \wb_i \in \Db,
\end{equation}
	where $\boldsymbol{e}_j$ is a unit vector in $\mathds{R}^{d+1}$, having $1$ in its $i$th element and zero otherwise. Using KKT conditions for solving the above problem, we have:
	\begin{subequations}
		\begin{align}
		\qquad &e_j+\lambda\,\Sigmab_k^{-1}(\wb_i-\mub_k)=0,\\
		 \qquad &(\wb_i-\mub_k)^T\Sigmab_k^{-1}(\wb_i-\mub_k)\leq Q_{d+1}(\gamma),\label{eq:appendix:KKT:bound}\\
		 \qquad &\lambda\Big((\wb_i-\mub_k)^T\Sigmab_k^{-1}(\wb_i-\mub_k)- Q_{d+1}(\gamma)\Big)=0,		\quad \lambda\geq 0. 
		\end{align}
	\end{subequations}
	The inequality in (\ref{eq:appendix:KKT:bound}) is tight. With some algebraic manipulation, we have 	$\wb_i-\mu_k(j)=\frac{1}{\lambda}\Sigmab \boldsymbol{e}_j$, and so, $\frac{1}{\lambda^2} \eb_j^T\Sigmab_k \Sigmab_k^{-1}\Sigmab_k\eb_j=Q_{d+1}(\gamma)$. Therefore
		$\wb_i=\pm \sqrt{\frac{Q_{d+1}(\gamma)}{\boldsymbol{e}_j^T\,\Sigmab_k\,\boldsymbol{e}_j}}\Sigmab_k \boldsymbol{e}_j+\mub_k,$ 
$\lambda =\pm\sqrt{\frac{\eb_j^T\Sigmab_k\eb_j}{Q_{d+1}(\gamma)}},$ and
        $\eb_j^T \wb_i=\pm \sqrt{Q_{d+1}(\gamma) {\eb_j^T\Sigmab_k\eb_j}}+\mu_k(j)$.
	
	If $\lambda$ is positive, we can find the maximum which is $+ \sqrt{Q_{d+1}(\gamma) {\eb_j^T\Sigmab_k\eb_j}}+\mu_k(j)$, and if $\lambda$ is negative, we can find the minimum which is $- \sqrt{Q_{d+1}(\gamma) {\eb_j^T\Sigmab_k\eb_j}}+\mu_k(j)$.

	Here, $\mu_k(j)$ is the $j$th element of $\mub_k$. If we set $j=d+1$, then the delay perception of user $j$ is in the following range:
	\begin{equation}
	-\sqrt{Q_{d+1}(\gamma) \boldsymbol{e}_{d+1}^T\Sigmab_k \boldsymbol{e}_{d+1}}<	\beta_i(t)-\mu_k(d+1)<\sqrt{Q_{d+1}(\gamma) \boldsymbol{e}_{d+1}^T\Sigmab_k \boldsymbol{e}_{d+1}},
	\end{equation}
	at least with probability $\gamma$. Hence, Theorem \ref{lemma:confidence region single mode} is proved. 
    \section{Proof of Theorem \ref{prop:packet length}}
		Since the queuing delay is much smaller than the duration of each time slot, we can assume that each packet arriving at a specific time slot will be served at the same time slot. For analyzing the packet delay, we consider a packet that just arrives in the system in time slot $\tau_k$, and find  $\text{Pr}(D>D_i^{\max})$ for this packet. When this packet arrives, there are $m$ packets in the system. From lemma \ref{corol:exponential service time}, we know that the serving time will be an exponential random variable. Since the exponential distribution is memoryless, there is no distinction between a packet already in service and the other packets. Therefore, the waiting time for the packet that has just arrived is the summation of $m$ exponential distributions. Also, the transmission delay for this packet will be another exponential random variable. Hence, the delay of a packet which arrives at time slot $\tau_k$ while there are $m$ packets in the system can be written as:
		\begin{equation}
		d(\tau_k,m)=t_s+t_1(\tau_k)+t_2(\tau_k)+\cdots+t_{m-1}(\tau_k)+t_c(\tau_k),
		\end{equation}
		where $t_i(\tau_k)$ is the service time for packet $i$ in the queue, and $t_c(\tau_k)$ is the service time for packet already in service. Also, $t_s$ is the service time for the packet that has just arrived.
		we seek to find $\text{Pr}(d(\tau_k,m)>D_i^\textrm{max})$ which can be written as
		\begin{align}
		\text{Pr}\big(d(\tau_k,m)>D_i^\textrm{max}\big)&=\sum_{m,k} \text{Pr}(D>D_i^{\textrm{max}}|m,\tau_k)\text{Pr}(m,\tau_k)\nonumber\\
		&=\sum_{m,k} \text{Pr}(D>D_i^{\textrm{max}}|m,\tau_k)\text{Pr}(m|\tau_k) \text{Pr}(\tau_k).
		\end{align}
		The probability that there are $m$ users in an M/M/1 queue at time slot $\tau_k$, i.e.  $\text{Pr}(m|\tau_k)$, can be written as (see \cite{papoulis2002probability}):
		$
		\text{Pr}(m|\tau_k)=\left(\frac{a_i(\tau_k)}{r_i(\tau_k)}\right)^m \left(1-\frac{a_i(\tau_k)}{r_i(\tau_k)}\right).
	$
		Since we assumed the time slots have equal lengths, the packets arrive at each time slot with equal probability of $\text{Pr}(\tau_k)=\frac{1}{t}$, where $t$ is the total number of time slots.
		
		The sum of $m+1$ identically independent exponential random variables with the mean $\frac{1}{r_i(\tau_k)}$ is a gamma random variable. Consequently, if the users arrive at time slot $\tau_k$ while there are $m$ users in the system at the time of arrival, the distribution of delay is
		\begin{equation}
		f_D(\phi|m,\tau_k)=\frac{r_i(\tau_k)^{m+1}}{\Gamma(m+1)}\phi^{m}e^{-r_i(\tau_k)\phi}.
		\end{equation}
		
		As a result, we can write the probability of delay exceeding a threshold $D_i^{\max}$ as
		\begin{align}
		\text{Pr}(D>D_i^{\max})=
		&\int_{D_i^{\max}}^{\infty} \sum_{m,k} f_D(\phi|m,\tau_k)\text{Pr}(m|\tau_k) \text{Pr}(\tau_k) d \phi\\
		= &\int_{D_i^{\max}}^{\infty} \frac{1}{t}\sum_{m,k} \frac{r_i(\tau_k)^{m+1}}{m!}\phi^{m}e^{-r_i(\tau_k)\phi}(\frac{a_i(\tau_k)}{r_i(\tau_k)})^m (1-\frac{a_i(\tau_k)}{r_i(\tau_k)}) d \phi\\
		=&\frac{1}{t}\sum_{k=1}^{t}\int_{D_i^{\max}}^\infty (r_i(\tau_k)-a_i(\tau_k))e^{-r_i(\tau_k)\phi}\sum_{m=0}^{\infty}\frac{\big(\phi a_i(\tau_k)\big)^m}{m!}d \phi\\
		=&\frac{1}{t}\sum_{k=1}^{t}\int_{D_i^{\max}}^\infty (r_i(\tau_k)-a_i(\tau_k))e^{-\big(r_i(\tau_k)-a_i(\tau_k)\big)\phi}d \phi\\
		=&\frac{1}{t}\sum_{k=1}^{t}e^{-\big(r_i(\tau_k)-a_i(\tau_k)\big)D_i^{\max}},
		\end{align}
	which proves the theorem.
	\nocite{*}
    \def\baselinestretch{1}
	\bibliographystyle{IEEEtran}
	\bibliography{PaperBib.bib}

\end{document}